# SUPERHUMPS IN CATACLYSMIC BINARIES.

# XXIV. TWENTY MORE DWARF NOVAE

JOSEPH PATTERSON,[1,2] JOHN R. THORSTENSEN,[3] JONATHAN KEMP,[4,2]

DAVID R. SKILLMAN,[5] TONNY VANMUNSTER,[6] DAVID A. HARVEY,[7] ROBERT A. FRIED,[8]

LASSE JENSEN,[9] LEWIS M. COOK,[10] ROBERT REA,[11] BERTO MONARD,[12]

JENNIE MCCORMICK,[13] FRED VELTHUIS,[13] STAN WALKER,[14] BRIAN MARTIN,[15]

GREG BOLT,[16] ELENA PAVLENKO,[17] DARRAGH O'DONOGHUE,[18] JERRY GUNN,[19]

RUDOLF NOVÁK,[20] GIANLUCA MASI,[21] GORDON GARRADD,[22] NEIL BUTTERWORTH,[23]

THOMAS KRAJCI,[24] JERRY FOOTE,[25] & EDWARD BESHORE[26]

[1] Department of Astronomy, Columbia University, 550 West 120th Street, New York, NY 10027; jop@astro.columbia.edu

[2] Visiting Astronomer, Cerro Tololo Interamerican Observatory, National Optical Astronomy Observatories, which is operated by the Association of Universities for Research in Astronomy, Inc. (AURA) under cooperative agreement with the National Science Foundation

[3] Department of Physics and Astronomy, Dartmouth College, 6127 Wilder Laboratory, Hanover, NH 03755; thorstensen@dartmouth.edu

[4] Joint Astronomy Centre, 660 North A`ohōkū Place, Hilo, HI 96720; j.kemp@jach.hawaii.edu

[5] Center for Backyard Astrophysics (East), 9517 Washington Avenue, Laurel, MD 20723; dskillman@home.com

[6] Center for Backyard Astrophysics (Belgium), Walhostraat 1A, B–3401 Landen, Belgium; Tonny.Vanmunster@cbabelgium.com

[7] Center for Backyard Astrophysics (West), 1552 West Chapala Drive, Tucson, AZ 85704; dharvery@comsoft-telescope.com

[8] Center for Backyard Astrophysics (Flagstaff), Braeside Observatory, Post Office Box 906, Flagstaff, AZ 86002; captain@asu.edu

[9] Center for Backyard Astrophysics (Denmark), Søndervej 38, DK–8350 Hundslund, Denmark; teist@image.dk

[10] Center for Backyard Astrophysics (Concord), 1730 Helix Court, Concord, CA 94518; lcoo@yahoo.com

[11] Center for Backyard Astrophysics (Nelson), 8 Regent Lane, Richmond, Nelson, New Zealand; reamarsh@ihug.co.nz

[12] Center for Backyard Astrophysics (Pretoria), Post Office Box 70284, Die Wilgers 0041, Pretoria, South Africa; lagmonar@csir.co.za

[13] Center for Backyard Astrophysics (Pakuranga), Farm Cove Observatory, 2/24 Rapallo Place, Farm Cove, Pakuranga, Auckland, New Zealand; jenniem@stardome.org.nz, fco@xtra.co.nz

[14] Center for Backyard Astrophysics (Waiharara), Wharemaru Observatory, Post Office Box 13, Awanui 0552, New Zealand; astroman@xtra.co.nz

[15] King's University College, Department of Physics, 9125 50th Street, Edmonton, AB T5H 2M1, Canada; bmartin@kingsu.ab.ca

[16] Center for Backyard Astrophysics (Perth), 295 Camberwarra Drive, Craigie, Western Australia 6025, Australia; gbolt@iinet.net.au

[17] Crimean Astrophysical Observatory, P/O Nauchny, 334413 Crimea, Ukraine; pavlenko@crao.crimea.ua

[18] South African Astronomical Observatory, Post Office Box 9, Observatory 7935, Cape Town, South Africa; dod@saao.ac.za

[19] Center for Backyard Astrophysics (Illinois), 1269 North Skyview Drive, East Peoria, IL 61611; jgunn@mtco.com

[20] Nicholas Copernicus Observatory, Kravi Hora 2, Brno 616 00, Czech Republic; novak@hvezdarna.cz

[21] Center for Backyard Astrophysics (Italy), Via Madonna de Loco, 47, 03023 Ceccano FR, Italy; gianluca@bellatrixobservatory.org

[22] Center for Backyard Astrophysics (Tamworth), Post Office Box 157, Tamworth NSW 2340, Australia; loomberah@ozemail.com.au

[23] Center for Backyard Astrophysics (Townsville), 24 Payne Street, Mount Louisa, Queensland 4814, Australia; neilbutt@bigpond.com.au

[24] Center for Backyard Astrophysics (New Mexico), 1688 Cross Bow Circle Drive, Clovis, NM 88101; krajcit@3lefties.com

[25] Center for Backyard Astrophysics (Utah), 4175 East Red Cliffs Drive, Kanab, UT 84741; jfoote@scopecraft.com

[26] Center for Backyard Astrophysics (Colorado), 14795 East Coachman Drive, Colorado Springs, CO 80908; ebeshore@pointsource.com

## ABSTRACT

We report precise measures of the orbital and superhump period in twenty more dwarf novae. For ten stars, we report new and confirmed spectroscopic periods — signifying the orbital period $P_o$ — as well as the superhump period $P_{sh}$. These are GX Cas, HO Del, HS Vir, BC UMa, RZ Leo, KV Dra, KS UMa, TU Crt, QW Ser, and RZ Sge. For the remaining ten, we report a medley of $P_o$ and $P_{sh}$ measurements from photometry; most are new, with some confirmations of previous values. These are KV And, LL And, WX Cet, MM Hya, AO Oct, V2051 Oph, NY Ser, KK Tel, HV Vir, and RX J1155.4–5641.

Periods, as usual, can be measured to high accuracy, and these are of special interest since they carry dynamical information about the binary. We still have not quite learned how to read the music, but a few things are clear. The fractional superhump excess ε $[=(P_{sh}-P_o)/P_o]$ varies smoothly with $P_o$. The scatter of the points about that smooth curve is quite low, and can be used to limit the intrinsic scatter in $M_1$, the white dwarf mass, and the mass-radius relation of the secondary. The dispersion in $M_1$ does not exceed 24%, and the secondary-star radii scatter by no more than 11% from a fixed mass-radius relation. For the well-behaved part of ε($P_o$) space, we estimate from superhump theory that the secondaries are 18±6% larger than theoretical ZAMS stars. This affects some other testable predictions about the secondaries: at a fixed $P_o$, it suggests that the secondaries are (compared with ZAMS predictions) 40±14% less massive, 12±4% smaller, 19±6% cooler, and less luminous by a factor 2.5(7). The presence of a well-defined mass-radius relation, reflected in a well-defined ε($P_o$) relation, strongly limits effects of nuclear evolution in the secondaries.

*Subject headings*: accretion, accretion disks — binaries: close — novae, cataclysmic variables — stars: individual (GX Cassiopeiae) — stars: individual (HO Delphini) — stars: individual (HV Virginis) — stars: individual (BC Ursae Majoris) — stars: individual (RZ Leonis) — stars: individual (KV Dracnonis) — stars: individual (KS Ursae Majoris) — stars: individual (TU Crateris) — stars: individual (QW Serpentis) — stars: individual (RZ Sagittae) — stars: individual (KV Andromedae) — stars: individual (LL Andromedae) — stars: individual (WX Ceti) — stars: individual (MM Hydrae) — stars: individual (AO Octantis) — stars: individual (V2051 Ophiuchi) — stars: individual (NY Serpentis) — stars: individual (KK Telescopii) — stars: individual (HV Virginis) — stars: individual (RX J1155.4–5641)

## 1. INTRODUCTION

*Superhumps* in the light curves of cataclysmic variable stars (CVs) are large-amplitude waves at a period slightly displaced from the orbital period $P_o$. The most common type occur at a period longer than $P_o$, and are now understood as arising from an eccentric instability at the 3:1 resonance in the accretion disk (Whitehurst 1988, Hirose & Osaki 1990, Lubow 1991, Murray 1998, Wood et al. 2000). The secondary star tugs on the eccentric disk and forces precession, and the superhump frequency $\omega_{sh}$ is then interpreted as the lower precessional sideband of the orbital frequency $\omega_o$. Most studies have concentrated on dwarf novae, because observations demonstrate a consistent phenomenology: superhumps are made in every superoutburst of SU Ursae Majoris-type dwarf novae, never in short outbursts, and never in any other type of dwarf nova. Good discussions are given by Osaki (1996), Hellier (2000, Chapter 6) and Warner (1995, Chapter 4). A one-page summary is given in Appendix A of Patterson et al. (2002).

A useful quantity, readily provided by observation, is the superhump's fractional period excess $\varepsilon$ [$\equiv(P_{sh}-P_o)/P_o$]. According to theory (Hirose & Osaki 1990) and observation (Stolz & Schoembs 1984; Patterson 1998, hereafter P98), $\varepsilon$ scales with $P_o$, basically because $P_o$ determines the secondary-star mass $M_2$, which supplies the perturbation causing the superhump. If we understood that scaling, we could use $\varepsilon$ to learn the mass ratio $q$, which is a key ingredient in understanding CV evolution but is difficult to constrain.

For several years we have been trying to advance this enterprise with precise independent measures of $P_o$ and $P_{sh}$ in dwarf novae (Thorstensen et al. 1996, Thorstensen & Taylor 1997, and previous papers in this superhump series). Here we present these measures for twenty additional dwarf novae, and show that the empirical $\varepsilon(P_o)$ relation accurately constrains the mass-radius relation for CV secondaries. This is critical for understanding why secondaries are less massive and cooler than expected under a ZAMS assumption — and for understanding why the minimum $P_o$ is as long as ~80 minutes.

## 2. SPECTROSCOPIC OBSERVATIONS

### 2.1 TECHNIQUES

Our spectroscopy is all from the 2.4 m Hiltner telescope at MDM Observatory. The instrumentation and techniques were generally as described by Thorstensen et al. (1998). The modular spectrograph, a 600 l/mm grating, and various CCD detectors typically gave a dispersion of 2.0 Å/pixel and a spectral resolution ~3.5 Å FWHM. For most of the observations a $2048^2$ CCD was used, providing coverage from around 4300 to 7500 Å, with vignetting near the end of the range. Some of the early observations were taken with a $1024^2$ CCD which covered from shortward of H$\beta$ to longward of H$\alpha$. Throughout the observations we maintained a wavelength calibration with frequent comparison lamp exposures, especially when the telescope was moved. The [O I] $\lambda$5577 night-sky line was typically stable in the background spectra to ~10 km/s. We observed bright B stars to calibrate telluric absorption features, flux standards when the sky appeared clear, and used these to convert our raw spectra to flux units. The slit was only 1 arcsec wide, so seeing and transparency fluctuations severely compromise the absolute flux scale, which we estimate is accurate only to ~30%. Fluxes for our program stars are more likely to be underestimated than overestimated, since standards were observed only in good conditions and were accurately centered. The final spectra sometimes showed unphysical wavelike variations with amplitude 10–20 percent, which we still don't understand; but these variations appear to average out over many exposures. Exposures were kept shorter than 480 s to minimize smearing in orbital phase.

We reduced the data with standard IRAF routines, and measured radial velocities of the Hα emission line using tunable convolution algorithms (Schneider & Young 1980). To search for periods we used a residual-gram algorithm (Thorstensen et al. 1986). Monte Carlo tests (Thorstensen & Freed 1985) were used as appropriate to assign likelihoods to various cycle-count aliases. We also prepared phase-averaged single-trailed representations of our spectra, using procedures described by Taylor, Thorstensen, & Patterson (1999).

In five of the ten stars studied, the spectroscopy did not unambiguously specify the daily cycle count. In two of these, the (narrowly) preferred cycle count was confirmed by the unambiguous cycle count in the photometry. In the other three (GX Cas, HO Del, RZ Sge) we rely on the assumption of approximate "normality" ($P_{sh}>P_o$) to decide the cycle count.

## *2.2 RESULTS*

We present our main results in tables and figures. Table 1 contains a journal of spectroscopy. Figure 1 shows the mean quiescent spectra, and Table 2 contains measurements of the emission lines. The FWHM measures in Table 2 are from Gaussian fits; the lines were often highly non-Gaussian, but the FWHM measures appeared approximately correct in most cases. The line fluxes are subject to the uncertainties mentioned earlier. Table 3 (which appears in its entirety in the electronic version) gives the individual velocity measurements. Figures 2 and 3 give the periodograms and folded velocity curves, and Table 4 gives parameters of the sinusoidal velocity fits. In four stars, we were able to trace faint "*S*-waves" in the single-trailed spectrum display. These were clearest in He I (λ5876 or λ6678), but could also be discerned in Hα. Figure 4 shows portions of the single-trailed spectra. We measured these single-trailed velocities versus phase by eye, using a cursor. The measured parameters are given in Table 5; but we did not use the velocities for any subsequent analysis, since the interpretation of *S*-waves is fraught with uncertainty (Smak 1985).

## 3. PHOTOMETRY AND PERIOD ANALYSIS

In brief, we measure photometric periods by splicing long light curves of differential magnitudes (with respect to a nearby comparison star), and calculating their power spectra. We used the Center for Backyard Astrophysics (CBA) telescope network. More details concerning the CBA can be found at http://cba.phys.columbia.edu/, and more details on data analysis are given by Skillman & Patterson (1993).

There are numerous fine details to the enterprise. In practice, photometry from one longitude is seldom sufficient to eliminate aliasing in frequency. This is partly because dwarf novae can be afflicted with large random variability, and partly because observations at large airmass have differential extinction effects which cannot be removed (especially with unfiltered data). So we always splice the light curve with contributions from observers over a range of longitudes. We also usually subtract the mean and (linear) trend from each night's time series. This is important, because it greatly reduces the low-frequency noise arising from erratic night-to-night variability.[1] We then examine the spliced light curve and look for obvious amplitude changes, which can corrupt power-spectrum analysis if they are severe. When there are strong amplitude changes (greater than a factor ~3), we eliminate the offending observation or re-analyze the time series in flux units (if that helps).

We present mainly "dirty" power spectra, in order to permit the reader to evaluate the quality of the evidence for the periodic signal and its discrimination of aliases. In a few cases we "clean" the power spectrum (correct for windowing of the time series), usually to study the presence of fine structure. We sometimes use period units, since humans tend to think that way, but more commonly frequency, since that makes transparent the alias problems that are a frequent hazard of this enterprise. To minimize the verbal morass, we generally use the old-style "c/d" notation, rather than cycles day$^{-1}$ as mandated by the local captains of correctness.

There is also an issue of accuracy. To be useful in the present context, we need to know the fractional period excess to an accuracy of ~20%. Frequencies can roughly be measured to an accuracy ~0.12/N c/d, where N is the baseline in days; so we require a baseline of at least 3 days to achieve the needed accuracy. In principle, the harmonics can improve the accuracy by a factor of 2 or 3; but only a small improvement is realistic, since the harmonics tend to be weak and somewhat variable. Additional uncertainty of ~0.02–0.03 c/d is often contributed by actual period changes during outburst (Figure 11 of Patterson et al. 1993, Figure 14 of Kato et al. 2003), or slightly less in case the period change can be adequately measured. In practice, the combined effective uncertainty in superhump frequency $\omega_{sh}$ is usually no better than 0.02–0.03 c/d. When this is further degraded by inadequate or aliased coverage, we reject the star.

## 4. INDIVIDUAL STARS, FROM PHOTOMETRY AND SPECTROSCOPY

Stars were selected for a coordinated CBA photometry campaign when reports from visual observers (usually) indicated a superoutburst. Each star performed the usual hijinks for SU UMa-type dwarf novae: rapid rise to maximum light; rapid onset of large-amplitude *common superhumps* at or near maximum; inital decay from maximum at ~0.13 mag/day; final rapid decline at ~1 mag/day after 10–20 days; cessation of superhumps during or after rapid decline. Of course, our photometry does not cover all phases of each superoutburst studied; but it is sufficiently extensive to show consistency with this "standard eruption". We omit the usual panoply of light curves, since common superhumps look practically the same in every dwarf nova. In a later publication we will discuss finer details for some of these stars.

Details of the photometric coverage are given in Table 6. The period searches are mainly shown in Figures 2 and 3 (spectroscopy) and 5 (photometry). Now we briefly summarize results for each of the 10 stars with spectroscopy, alphabetically by constellation.

---

[1] Of course, it also eliminates *true* signals at low frequency! So we do not use this when we want to study very low frequencies.

### *4.1 GX CASSIOPEIAE*

Photometry of GX Cas was obtained for 9 days during supermaximum in October 1995, giving a peak in Figure 5 at 10.75(3) c/d. This agrees with the value given by Nogami, Kato, & Masuda (1998). The object proved difficult spectroscopically, with velocities yielding an ambiguous daily cycle count, but the unambiguous choice from (superhump) photometry indicates that the favored $P_o$ is the correct one. The double peaks of the H$\alpha$ line are separated by 1000 km/s. Liu & Hu (2000) present a spectrum which appears consistent with ours. An *S*-wave is present in the single-trailed representation of Figure 4.

### *4.2 TU CRATERIS = J 05.23*

TU Crt was discovered by Maza et al. (1992) and studied further by Koen & O'Donoghue (1992). Photometry was obtained over 12 days in the February 1998 superoutburst, giving a signal at 11.716(24) c/d. Details of this CBA campaign were reported earlier (Mennickent et al. 1999a), and so the star is omitted from Figure 5. The H$\alpha$ velocities yielded an unambiguous solution for $\omega_o$ at 12.182(13) c/d. The continuum is blue, with a weak depression around H$\beta$ which might indicate a substantial contribution from a white-dwarf photosphere. As noted by Maza et al. (1992), the double peaks of H$\alpha$ are separated by about 920 km/s. An *S*-wave is just discernible in He I $\lambda$6678 (Figure 4).

### *4.3 HO DELPHINI*

Superhumps of HO Del were observed for 5 days in September 1996, giving a peak in Figure 5 at $\omega_{sh}$=15.54(4) c/d. Our spectroscopy at quiescence ($V\sim19$) gave the strongest peak in Figure 2 at $\omega_o$=15.96(4) c/d; the latter has some possibility of cycle count error, but that appears to be excluded by the superhump photometry. Munari and Zwitter (1998) give a minimum-light spectrum which appears generally similar to ours, but with higher flux.

### *4.4 KV DRACONIS = HS1449+6415*

KV Dra was first identified as an X-ray-emitting CV in the Hamburg-Rosat objective prism survey (Nogami et al. 2000). Vanmunster et al. (2000) gave a preliminary account of the CBA campaign during the May 2000 superoutburst. The 12-day coverage enabled a fairly good estimate of the mean superhump frequency, shown in Figure 5 as 16.63(3) c/d. The fine details showed strong amplitude variations, which tend to pollute the power spectrum; but the mean frequency was fairly well determined. The radial-velocity search was normal and gave an unambiguous $\omega_o$=17.02(2) c/d. Velocities were taken on two observing runs spaced by 86 days; combining these data sets constrains the period to 85.940(3) d / $N$, where $1457 \leq N \leq 1471$. The period at minimum found by Nogami et al. (2000) corresponds to $N$=1457. Our best spectroscopic period differs from this by about 3$\sigma$, and corresponds to $N$=1463.

### *4.5 RZ LEONIS*

One of the very rare eruptions of RZ Leo was observed in December 2000. Ten days of coverage gave $P_{sh}$=0.07868(18) d, as seen in Figure 5. A good discussion of RZ Leo's eruptions is given by Ishioka et al. (2001), who also provide more detail about this particular eruption.

Previous study in quiescence has shown a photometric signal with period 0.0708(3) and 0.0756(12) d (respectively Howell & Szkody 1988 and Mennickent et al. 1999b). This discrepancy warranted reobservation, so we obtained quiescent photometry in 1998–9. A densely spaced four-night run gave the power spectrum seen in the upper frame of Figure 6, revealing a frequency of 13.15(2) c/d and a double-humped waveform. Additional observations over the next year confirmed the stability of the frequency and waveform. The lower frame of Figure 6 shows the mean light curve, and an O–C diagram illustrating the stability of the timings of maximum light, with respect to the orbital ephemeris:

$$\text{Primary maximum} = \text{HJD } 2,450,851.0477(14) + 0.0760383(7) \, E \, . \tag{1}$$

Previous spectroscopy at quiescence has yielded $P_o$=0.07651(26) d (Mennickent & Tappert 2001). This agrees with our spectroscopic period of 0.0761(2) d. The Balmer profiles are strongly double-peaked, with a peak separation of almost 1300 km/s. The system is faint and difficult, and much of the "velocity" signal arises from variation in the ratio of the violet and red peaks (*V/R* variation). The single-trailed spectrum in Figure 4 shows a clear *S*-wave in He I λ5876, further corroborating the orbital period which was used to construct the image. The *S*-wave's large amplitude, along with the line doubling and large peak separation, indicates that the binary inclination must be fairly high.

*4.6 RZ SAGITTAE*

Superhumps of RZ Sge were observed for 5 days during the August 1996 supermaximum, yielding the peak seen in Figure 5 at 14.17(3) c/d. There are several previous detections (Bond, Kemper, & Mattei 1982; Kato 1996; Semeniuk et al. 1997), yielding signals respectively at 14.26, 14.20, and 14.21 c/d. We average these and adopt 14.21(3) c/d, or $P_{sh}$=0.07037(15) d.

Because of observing season and hour-angle constraints, our spectroscopy at quiescence did not resolve between candidates at 13.65 and 14.65 c/d; we rely on the superhumps to establish a preference (i.e., select the candidate which gives $P_{sh}>P_o$). Our June 1999 run then selected a period of 0.06828(2) d. For future reference, the precise $P_o$ is 273.122(3)/*N* d, where *N*=4000±4 is an integer. The Hα line profile is double-peaked, with peaks separated by 780 km/s. A nascent wave appeared in the single-trailed spectrum, but not clearly enough to measure.

*4.7 QW SERPENTIS = TkV46*

Superhumps in QW Ser were observed for 6 days in July 2000, yielding $P_{sh}$= 0.07698(23) d. Our spectroscopy is not very extensive, but it does establish a unique $P_o$=0.07453(10) d. In the average spectrum, Hα is double-peaked with a separation of 1100 km/s, so the orbital inclination is probably fairly high.

### *4.8 BC URSAE MAJORIS*

Like RZ Leo, this is another borderline WZ Sge star — a dwarf nova with very few normal outbursts and a long recurrence period (~1000 d between supermaxima). We observed a long outburst in February 2000, and found strong superhumps with $\omega_{sh}$=15.50(3) c/d, illustrated in Figure 5. More detail on the eruption is given in Figure 7. On the first 4 nights (JD 2451635–8) the light curves were very quiet, with the power spectrum (upper frame of Figure 7) showing a complex near 32 c/d. The highest peak occurred at 31.97(4) c/d, consistent with $2\omega_o$ as deduced below. Synchronous summation at $\omega_o$ gave a low-amplitude double sinusoid, inset in Figure 7. This appears to be a manifestation of an "outburst orbital hump" or "early superhump", a common signature of WZ Sge-type dwarf novae. The origin of these transient waves is still unknown, although extensively discussed in the context of WZ Sge's recent eruption (Osaki & Meyer 2002, Kato 2002, Patterson et al. 2002). On JD 2451639 common superhumps grew rapidly (lower frame of Figure 7), and, as usual, dominated the light curve for the rest of the eruption.

In February 1999 we acquired a 40-day observation in or near quiescence, including a dense segment lasting 6 days. The power spectrum of the latter is shown in Figure 8, with a signal at 31.92(5) c/d. The full 40-day observation was afflicted by aliasing, with equal peaks at 31.950 and 31.978 (±0.003) c/d, and weak signals at 15.941, 15.970, and 15.999 (±0.003) c/d. Looking for solutions in a 1:2 ratio, the favored choice is at $\omega_o$=15.973(4) c/d. A more conservative choice, free from aliasing, is 15.97(2) c/d. The mean quiescent light curve is inset in Figure 8; the double hump is a characteristic feature of these low-$\dot{M}$ dwarf novae.

The spectrum shows broad absorption around H$\beta$ and a blue continuum, which indicates a strong white dwarf contribution. Previous spectra obtained by Mukai et al. (1990) and Smith et al. (1997) also show this, as well as a secondary star contribution of type M5+ in the near infrared. The radial-velocity search in Figure 3 gave two acceptable (aliased) values of $\omega_o$, but the photometric period unambiguously selects the correct choice.

### *4.9 KS URSAE MAJORIS = SBS 1017+533*

This star was discovered in the Second Byurakan Sky Survey (Markarian & Stepanian 1983), and identified as a CV by Balayan (1997). Eruptions to *V*~12.5 have been found over the last century (Hazen & Garnavich 1999), and recent intensive coverage by visual observers have shown the eruptions to have the "long and short" dichotomy characteristic of SU UMa-type dwarf novae.

Photometry was obtained during the Feb 1998 and May 1999 superoutbursts. The 1998 eruption showed a strong signal at 14.32(4) c/d. The 1999 eruption was covered longer (12 days), but the sampling was less favorable. The first 6 days of the eruption showed fairly stable superhumps, shown in Figure 5 at 14.42(3) c/d. After day 10, the beginning of the rapid decline phase, the superhumps became double-humped and then mutated into "late" superhumps; it is possible that this behavior started a little earlier, contributing to the noise in Figure 5.

This difference in $\omega_{sh}$ between well-observed superoutbursts is unusual — but not so unusual as to warrant an explanation! As a compromise we adopt $\omega_{sh}$=14.37(4) c/d.

Radial velocities favor $P_o$=0.06796(10) d, with some concern about aliasing as seen in Figure 3. We resolve this uncertainty by using the unambiguous $P_{sh}$. The emission lines are fairly narrow, and the single-trailed spectrum is uninformative, suggesting a rather low binary inclination. Jiang et al. (2000) also show a spectrum of this object which resembles the one presented here.

### *4.10 HS VIRGINIS*

Photometry of HS Vir was obtained during the March 1996 superoutburst. Most of our data is in the March 16–20 window, about 9 days after the eruption actually started. The power spectrum in Figure 5 is dominated by a large peak at 25.02 c/d. Kato et al. (1998) reported coverage of early phases of the same outburst, and found $\omega_{sh}$=12.41 c/d. The superhump evidently developed a double-humped structure, and decreased slightly in period. These changes are common in dwarf novae, and indeed were found for this particular eruption by Kato et al. We adopt 12.43(3) c/d, or $P_{sh}$=0.08045(19) d.

Mennickent, Matsumoto, & Arenas (1999) reported quiescent spectroscopy which gave $P_o$=0.07692(3) d, with some aliasing worries. Our value of 0.0769(2) d confirms their $P_o$ and gives some extra weight to the cycle count leading to their preferred $P_o$. We extracted the published velocities from their Table 3 and analyzed them jointly with ours; because the two data sets are two years apart, there is ambiguity in cycle count between the observing runs. The precise periods corresponding to Mennickent et al.'s preferred $P_o$ are given by $P_o$=754.1358 d/$N$, where $N$ is an integer between 9802 and 9807. The corresponding figures for their two alternate choices are (754.0833 d, 9784<$N$<9788) and (754.1111 d, 9819<$N$<9824).

## 5. INDIVIDUAL STARS, FROM PHOTOMETRY

For the ten stars described above, we have confirmed spectroscopic and superhump periods in our own data. For the remaining ten we do not have new spectroscopy, but nevertheless have enough data to establish or confirm unique values of $P_o$ and $P_{sh}$ from photometry alone (our own or published). Now we discuss those ten.

### *5.1 LL ANDROMEDAE*

LL And is a little-studied dwarf nova, erupting at intervals of a few years. One such eruption was observed in 1994 to show superhumps with $P_{sh}$=0.0567 d (Table 7 of Kato et al. 2001). We eagerly await a chance to observe these humps; but the star has proved mighty bashful. Quiescent photometry in December 1997 did, however, reveal a strictly periodic signal very likely to be $P_o$. Figure 9 shows the power spectrum and mean light curve over the 30-day baseline. The orbital ephemeris is

$$\text{Minimum light} = \text{HJD } 2450787.6189(10) + 0.055055(6) \, E \, . \qquad (2)$$

The origin of the orbital signal is not known, of course; but such things are commonly seen in quiescent dwarf novae. This implies $\varepsilon=0.030$ (depending on the unknown error in $P_{sh}$), a normal value which requires a fairly normal secondary near 0.1 $M_\odot$.

### 5.2 KV ANDROMEDAE

In January 1999 we obtained a 9-day time series on KV And in superoutburst. A garden-variety superhump was observed, at $\omega_{sh}=13.45(2)$ c/d as seen in the upper frame of Figure 10. This agrees with the result of Kato (1995). In October 1999 we obtained a 4-day time series near quiescence ($V\sim19.5$), and found a powerful signal at 13.65(4) c/d. This is seen, along with the corresponding mean light curve, in the lower frame of Figure 10. Interpreted as $\omega_o$, this implies $\varepsilon=0.0145(32)$, an interestingly low value at this orbital period.

Superhumps can linger a very long time after superoutburst, even approaching quiescence; and this star is rather sparsely observed. Therefore it is possible that the "quiescent" signal is merely the late-superhump residue of an unobserved outburst. We searched variable-star records to evaluate this possibility, with inconclusive results. The star is included here for completeness; we consider the $P_o$ value as likely, but in need of confirmation from spectroscopy.

### 5.3 WX CETI

A thorough study of WX Ceti's superhumps was presented by Kato et al. (2001), who found a signal at 16.810(3) c/d in a November 1998 superoutburst. We obtained a 12-night time series during the same outburst, and found a signal at $\omega_{sh}=16.80(2)$ c/d, seen in Figure 11. Another superoutburst was observed in July/August 1996, and 6-day time series gave $\omega_{sh}=16.84(2)$. We average these to estimate $\omega_{sh}=16.81(2)$. Thorstensen et al. (1996) measured $\omega_o=17.164(4)$ c/d from radial velocities at quiescence.

Most of the 1996 data was high-speed (3 s time resolution) photoelectric photometry, suitable for periodicity search at very high frequencies. The first five days of observation gave no detection (to a semi-amplitude upper limit of 0.0006 mag), but the sixth day showed a strong signal at 17.385(6) s. This signal of semi-amplitude 0.0015 mag is seen in the power spectrum of Figure 12. It is evidently a "dwarf nova oscillation", a fairly common syndrome of dwarf novae at maximum light (Patterson 1981).

### 5.4 MM HYDRAE = PG 0911–066

Misselt & Shafter (1995) found large-amplitude waves, resembling superhumps, in this star. After a year's close watch, we were rewarded in March 1998 when the star jumped to $V=13.8$ and showed obvious superhumps. Five days of dense coverage showed the power spectrum in the upper frame of Figure 13, yielding $\omega_{sh}=17.04(2)$ c/d. Thirty days later, the star had declined to quiescence at $V\sim18.5$, and we obtained a 12-night time series. The corresponding power spectrum is shown in the second frame of Figure 13. The low-frequency region is complex, but at higher frequency the signal is a simple (windowed) sinusoid at 34.722 c/d. The latter is the key to understanding the structure at low frequency — for the lightly cleaned version of the quiescent power spectrum, seen in the third frame, shows two independent frequencies: the residue of the superhump at 17.09(1) c/d, and a signal at 17.367(6) c/d with comparable power at the first harmonic.

The latter is almost certainly $\omega_o$. By combining with five other nights at quiescence over a 110-day baseline, we were able to refine $P_o$ and establish an ephemeris during 1997–8:

$$\text{Maximum light} = \text{HJD } 2451809.885(2) + 0.057590(2)\, E\,. \qquad (3)$$

The lowest frame of Figure 13 shows the O–C diagram of maxima relative to this ephemeris, and the mean orbital waveform.

### 5.5 AO OCTANTIS

We obtained a 5-night time series on this star during its September 2000 superoutburst. The power spectrum in the upper frame of Figure 14 shows a signal at 14.89(3) c/d. A 4-night observation in quiescence gave the signal in the lower frame, at 15.25(3) c/d. Inset is the mean quiescent light curve, which tracks

$$\text{Maximum light} = \text{HJD } 2451400.665(2) + 0.06557(13)\, E\,. \qquad (4)$$

### 5.6 V2051 OPHIUCHI

V2051 Ophiuchi is a CV of uncertain family linkage. Periodic deep eclipses establish $\omega_o$=16.019(<1) c/d. The star is most commonly near $V$=15 and erupts to $V$~13 on a timescale ~500 d; but the brightness levels and recurrence times are highly variable — perhaps more so than any SU UMa star of our acquaintance. Indeed, the star's credentials for assignment to the SU UMa class are far from solid. Vrielmann, Stiening, & Offutt (2002) found difficulty in a disk interpretation, and Warner & O'Donoghue (1987) interpreted the star as a low-field polar, despite the absence of the usual credentials for magnetism (polarimetry, X-rays, stable spin period). It does, however, show superhumps, so we include it here.

We obtained a 12-night time series in the May/June 1998 eruption, giving the power spectrum in the upper frame of Figure 15. The main signal occurred at 15.70(2) c/d, although close inspection showed a change from 15.60 to 15.71 (both ±0.04) c/d. The signal at higher frequency is puzzling; 31.08 c/d could be $2\omega_{sh}$, but only if the signal mainly occurs early in the outburst, contrary to the (somewhat weak) evidence. Kiyota & Kato (1998) measured $\omega_{sh}$=15.57 from a 3-night time series early in the same outburst, with an error we estimate as ±0.07 c/d.

Nine consecutive nights of good coverage were obtained in the Jul/Aug 1999 eruption, with the power spectrum seen in the lower frame of Figure 15. The fundamental is well-defined at 15.53(3) c/d. The forest of peaks at higher frequency arises from the presence of two signals, convolved with the spectral window. These were the (narrowly) highest peaks, at 31.27 and 32.04 c/d. The latter is obviously $2\omega_o$. The former is, strictly speaking, inconsistent with $2\omega_{sh}$;

but it could be consistent if the fundamental and the harmonic dominated in different parts of the outburst (with some migration in frequency). We could not conclusively prove this to be the correct explanation; but this is somewhat commonly seen in dwarf novae, because superhumps tend to evolve towards shorter period and greater secondary-hump structure.

From the above we estimate an average $\omega_{sh}$=15.58(4) c/d. These complexities are seldom seen in well-observed SU UMa stars; perhaps that is yet another aspect of the star's oddity.

### 5.7 NY SERPENTIS

This star was discovered as a UV-excess object in the Palomar-Green survey, and then identified as a dwarf nova by Nogami et al. (1998), who reported a superhump period of 0.106 d. We studied the star intensively, accumulating 270 hours over 57 nights, mostly in 1999. Short outbursts occur at 7 d intervals, and superoutbursts every 60–90 d. This affinity for eruption was problematic for the orbital period search, because erupting dwarf novae characteristically lose both emission lines and orbital modulations in photometry. Nevertheless, we managed to find an 8-day stretch suitable for period-finding. The star was in quiescence for the first 3 and the last 2 days, and in a short outburst for the middle 3. Like that of U Gem and many other dwarf novae, NY Ser's orbital modulation stays essentially constant in intensity units during outburst. So by converting to intensity units and subtracting the eruption light, we were able to track the orbital modulation throughout the 8 days. The resultant power spectrum, seen in the upper frame of Figure 16, shows an obvious signal at 10.26(2) c/d, or $P_o$=0.0975(2) d. This period is secure, and the inset waveform looks similar to those of many other dwarf novae in quiescence. The orbital wave tracked

$$\text{Maximum light} = \text{HJD } 2451255.875(5) + 0.09747(20)\, E \,. \qquad (5)$$

By using the full 90-day baseline of observations (and again trying to subtract eruption light in intensity units), we found a more accurate period, namely 0.09756(3) d. We don't have high confidence in this estimate, however, because of uncertainty in the subtraction.

A 5-day time series was obtained in a June 1999 superoutburst, giving an obvious signal at 9.63(3) c/d, seen in the lower frame of Figure 16.

### 5.8 KK TELESCOPII

A superoutburst of KK Tel was observed in May 2000, with a 6-day time series showing a signal (upper frame of Figure 17) at 11.41(2) c/d. A 5-night time series in quiescence in August 1999 (lower frame of Figure 17) gave a signal at 11.83(3) c/d.

### 5.9 HV VIRGINIS

This is another dwarf nova of very long recurrence period, with well-observed supermaxima in 1992 and 2002. Superhumps were found in the 1992 eruption (Barwig et al. 1992, Leibowitz et al. 1994), but with cycle count ambiguity, corrected by Kato, Sekine, & Hirata (2001). We obtained a dense 10-day time series in the 2002 outburst, producing the power spectrum seen in the upper frame of Figure 18. Inset is the mean light curve, a garden-variety superhump.

Photometry was obtained at quiescence during March-April 1998. The middle frame of Figure 18 shows the power spectrum of a dense 5-day time series, with a prominent signal at 35.02(3) c/d. This is evidently the signature of a double-humped wave at $\omega_o$. Also apparent is excess power near 11 c/d, including a possibly coherent signal at 11.20 or 12.21 c/d. We did not obtain enough data to assess the significance of the latter; some of the observing runs are relatively short — around 2–3 hours — so for this particular time series, we are inclined to distrust this feature on a similar timescale.

Quiescent photometry spanned a total of 58 days, and the lowest frame of Figure 18 shows the relevant portion of the power spectrum, with a signal at 35.045(4) c/d. At right is the mean light curve, tracking the orbital ephemeris

$$\text{Maximum light} = \text{HJD } 2450872.859(2) + 0.057069(6)\, E \,. \qquad (6)$$

The mean superhump periods from 2002 and 1992 are essentially identical at 0.05821(5) d, so the fractional period excess $\varepsilon = 0.0200(9)$.

### 5.10 RX J1155.4–5641

An eruption of this X-ray-selected dwarf nova was observed in April 2002. The observations extended for a total of 40 days, with an 8-day segment near maximum light particularly well sampled. The power spectrum of that dense segment is seen in the upper frame of Figure 19, showing strong signals at 16.07 and 32.21(3) c/d. We combine these to estimate $\omega_{sh}=16.09(2)$ c/d.

After subtracting the superhump from that light curve, we recomputed the power spectrum and found an obvious signal at 16.60(3) c/d. This seemed a reasonable candidate for $\omega_o$. Thirty days later, the star seemed to reach a fairly stable brightness level ~3.5 mag fainter. We interpreted this as quiescence. A power spectrum at that time contained a signal at 33.16(2) c/d, aliased by the poor sampling and shown in the lower right frame of Figure 19. This seemed a good candidate for $2\omega_o$, suggesting $\omega_o=16.59(2)$. This should be tested with spectroscopy or more extensive photometry at quiescence.

## 6. EPSILON VERSUS $P_o$

### 6.1 TABLE AND CAVEATS

Table 8 contains the basic ($P_o$, $P_{sh}$) data deduced in this work, along with four other period-pairs recently measured. The issue of errors, mentioned briefly in Sec. 3, deserves a second emphasis. For most dwarf novae, the minimum uncertainty in $P_{sh}$ is ~0.0002 d, because $P_{sh}$ can change by ~5 times that amount over a two-week outburst – and experience shows that we can only correct measurements to a "standard" phase in the outburst with an accuracy of ~0.0002 d. If the observations tend to cluster late in the outburst, the error grows since the humps can mutate into "late superhumps", with an attendant phase change corrupting the $P_{sh}$ measurement unless the photometry is sufficiently extensive to *prove* the phase change, which is rare. There are a few stars where the superhump is much more stable,[2] so we have tried to fine-tune the error estimates for each star accordingly.

Table 8 is the add-on to Table 4 of P98 and Table 3 of P01, and is the main goal of this research. Altogether there are 72 independent values of ε for apsidal superhumps of H-rich stars (with repeat values resolved in favor of the later tabulation). This includes ε in the "permanent" apsidal superhumps of novalike variables; these are not *exactly* like the common superhumps of dwarf novae, but the continuity in ε is noteworthy and a common or closely related origin is likely. The distribution of ε with $P_o$ is shown in Figure 20. We will see below, extending the discussion in P98 and P01, that $\varepsilon(P_o)$ provides a powerful diagnostic for fundamental binary parameters.

### 6.2 MAKING ε

A Roche-lobe filling secondary in a CV obeys the relation

$$P_o [\mathrm{hr}] = 8.75 \, (M_2/R_2^3)^{-\frac{1}{2}}, \tag{7}$$

with $M_2$ and $R_2$ in solar units (Faulkner, Flannery, & Warner 1972). We parameterize $R_2$ in terms of the "BCAH" radius $R_o$ for a star of solar composition at $t$=10 Gyr (Baraffe et al. 1998; such great ages[3] are needed to get the low-mass stars to shrink to a so-called "main sequence"). An approximation to the BCAH mass-radius relation, valid to 2% in the relevant 0.1–0.4 $M_\odot$ regime, is

$$R_o = 0.82 \, M_2^{0.82}, \tag{8}$$

so we write $R_2 = \alpha R_o$ to parameterize any departure from the main sequence. This yields

$$M_2 = 0.0764 \, P_o^{1.37} \, \alpha^{-2.05}. \tag{9}$$

Since this relation purports to give $M_2$ as a function of $P_o$, and $M_2$ is the perturbation that drives superhumps, we can use it to predict the run of ε with $P_o$. An empirical $\varepsilon(q)$ relation, based on 8 eclipsing CVs with $q$ sufficiently known to provide calibration, is $\varepsilon = 0.216(\pm 0.018)q$ (P01). We then expect ε to scale as

$$\varepsilon = 0.0165 \, P_o^{1.37} \, \alpha^{-2.05} \, M_1^{-1}, \tag{10}$$

---

[2] EG Cancri and WZ Sagittae are the prime examples, and are also the stars of lowest ε; thus this may well arise from the weakness of the perturbation ($M_2$) on the disk.

[3] We use this extreme assumption to get as low as 0.075 $M_\odot$, a regime interesting for CV evolution. For the more common 0.1–0.3 $M_\odot$ regime, the ZAMS is reached in <1 Gyr, so adopting such a great age does not place any great restriction.

where $M_1$ is also in solar units. The solid curve in Figure 20 shows the expected $\varepsilon(P_o)$ for BCAH secondaries and $M_1=0.75$, with the dashed extension representing the effect of losing thermal equilibrium at short $P_o$ (Kolb & Baraffe 1999, P01).

Obviously the data fall well below the predicted curve. Nevertheless, most of the points appear to define a curve which resembles the BCAH prediction, displaced downward, with remarkably little scatter. From the observed scatter in $\varepsilon(P_o)$ we can use superhump theory to measure dispersion in the relevant variables. Excluding the outliers at very short $P_o$, the rms scatter in $\varepsilon$ at a fixed $P_o$ is 21±3%. From Eq. (10), if this is entirely due to $M_1$, it implies a dispersion of 21±3% in $M_1$. If it is entirely due to $R_2$ (i.e., $\alpha$), then the dispersion in $\alpha$ is 11±2%. These are essentially *upper limits* to the dispersions, since the $\sigma$s add in quadrature (and since other variables, not included in this discussion, may contribute to the scatter at some level[4]).

### 6.3 $M_1$ VERSUS $R_2$

We can also use the actual values of $\varepsilon$ (rather than their dispersions) to limit the range of acceptable $<M_1>$ and $\alpha$ in the tidal theory of superhumps. This result depends on the adopted $\varepsilon(q)$ relation, but only for dwarf novae in the 0.06–0.085 d range, just where the calibration is most accurate — and where the stars are maximally homogeneous (SU UMa-type dwarf novae). There are 46 stars in this period range, and formally they each give an ($\alpha$, $M_1$) relation from Eq. (10). But the stars doubtless span a range in $\alpha$ and $M_1$, and superhump theory gives no way to decouple these variables. So instead we lump all 46 stars together, and characterize the average values of $\alpha$ and $M_1$.

The result is seen in Figure 21, where the shaded region indicates the ±2$\sigma$ error, including uncertainty in $\varepsilon(q)$ as well as errors in $\varepsilon$. To our wonderment and embarrassment, the curve goes right through (1, 1) — a crushing result after 10 years of research! What was the point of making all these measurements, anyway?! Is it possible that the cartoon CV of the 1970s ("1 $M_\odot$ white dwarf accreting from a ZAMS secondary") had it right after all?

Well... maybe *possible*, but very unlikely. In this range of $P_o$, there are six calibrating stars with a measured $M_1$ and an $\alpha$ deduced from (9). These are superimposed on Figure 21. A weighted average gives $<M_1>$=0.74(5), $<\alpha>$=1.24(8).[5] This value of $<M_1>$ is in good agreement with other published estimates for CVs as a class [0.74(4), Webbink 1990; 0.69(13), Smith & Dhillon 1997].

---

[4] For example, pressure effects in the disk (Lubow 1992; Ichikawa et al. 1993; Murray 2000). By using an empirical $\varepsilon(q)$ prescription, we have managed to eliminate the need to understand these. But they still exist, and should cause some of the scatter in Figure 20.

[5] The fact that the points are near the shaded curve is not of much significance, since these stars partly calibrate the adopted $\varepsilon(q)$ relation. Their slight (upward) displacement from the curve arises from the fact that the curve is defined by all 46 stars, not just these 6. The main point here is simply to identify a plausible choice of $<M_1>$ — thus enabling us to avoid the dreaded and embarrassing (1, 1) choice!

The conclusion is that superhump theory is satisfied if the secondaries have a radius averaging 18±6% greater then theoretical ZAMS stars. Does this mean that CV secondaries are "evolved"? No, that would be a great exaggeration. This is really a pretty small departure, considering the thousand natural shocks that CV secondaries are heir to (rapid rotation, struggles with thermal adjustment, a different outer boundary condition, eruptions from their jittery neighbors, etc.). It is also true that observations of single stars (and stars in wide binaries) do not yet test the BCAH mass-radius relation with adequate precision, in this low-mass regime; some fault could lie with the models.

Regardless of whether the BCAH models stand the test of time, the radii we derive still have the same value; we use BCAH merely as a benchmark. A little algebra shows how the other physical variables depend on $\alpha$:

$$M_2 = 0.0764 \, P_o^{1.37} \, \alpha^{-2.05} \quad (11)$$

$$L_2 = 0.0004 \, P_o^{3.70} \, \alpha^{-5.53} \quad (12)$$

$$R_2 = 0.0995 \, P_o^{1.12} \, \alpha^{-0.68} \quad (13)$$

$$T_2 = 2600 \text{ K } P_o^{0.36} \, \alpha^{-1.04}, \quad (14)$$

where all the quantities except $T_2$ are in solar units. Here we have adopted a power-law approximation to the BCAH results for mass-luminosity ($L_2 = 0.42 M_2^{2.70}$). The latter relationship should be more secure than mass-radius, since luminosity is well determined by central pressure and density (and hence mass).

With $\alpha = 1.18(6)$ deduced from the 46 "normal" dwarf novae, we find that at a fixed $P_o$, all of these variables are driven significantly lower. The expected mass falls by a factor 1.40(14); the radius falls by a factor 1.12(4); the temperature falls by a factor 1.19(6); and the luminosity falls by a factor 2.5(7).

All of these have significant consequences for CVs; but the latter two are perhaps the most noteworthy, since it has become popular to plot spectral type as a function of $P_o$. It has been known for a long time (Figure 4 of Patterson 1984, Figure 5 of Beuermann et al. 1998, Figure 3 of Baraffe 2002, Figure 5 of P01; also Friend et al. 1990) that CV secondaries are somewhat too cool, as compared to theoretical ZAMS stars. The reason is simple enough (as pointed out in 1984): the secondaries are too big! The fall in luminosity and temperature has a particularly severe effect on the detectability of secondaries at short $P_o$. At $P_o = 1.5$, $T_2$ is expected to be 2530 rather than 3010 K. Combined with the reduced luminosity, this implies that the secondary will be 3.0 mag fainter in V (1.8 in I, 1.2 in K). In other words, the (V, I, K) absolute magnitudes are expected to be (17.5, 13.4, 10.0). The white dwarfs and accretion light have typically $M_V = 10-12$ and neutral colors, so the secondaries are very elusive indeed.

*6.3 MASS–RADIUS, AND MINIMUM ORBITAL PERIOD*

Figure 21 is useful because it illustrates the connection between $\alpha$ and $<M_1>$ to satisfy the superhump data, in a well-populated and uncomplicated region of $P_o$ space. This result should be pretty secure. With somewhat less confidence, we could use $\varepsilon$ to infer $M_2$ [therefore $R_2$, via Eq. (7)] for each individual star — and thus assemble an empirical mass-radius relation. This is less accurate since it requires (for most CVs) an assumed $M_1$. The dispersion in $M_1$, estimated above, could be as high as 24%, which yields an 8% dispersion in $R_2$. Still, that is an estimate worth making.

We have done this earlier in Figure 2 of P01, and suggested an empirical formula

$$R_2 = 0.078 + 0.415\, M_2 + 3.16(M_2)^2 - 5.17(M_2)^3 \qquad [0.04 < M_2 < 0.36] \qquad (15)$$

(Eq. 7 of P01). Since there is no qualitative change in the data, we will not repeat it here. Basically, the data suggest radii ~15% greater than BCAH for $M_2 > 0.1\, M_\odot$, increasing to ~30% greater for 0.05–0.08 $M_\odot$.

Because the secondaries obey Eq. (7), these larger radii are in a sense the reason that the minimum $P_o$ is as long as 75–80 minutes, rather than ~70 minutes as generally predicted from theory (e.g. Paczynski 1981, Kolb & Baraffe 1999, Renvoize et al. 2002). We still do not understand why the secondaries are that large, however. An attractive possibility is that residual angular momentum loss (beyond that carried away by gravitational waves) continues throughout this short-period regime (P98). This would ameliorate several problems. It would drive the secondary further from thermal equilibrium and hence produce bigger $R_2$. It would increase the minimum $P_o$. It would shorten the lifetimes of CVs, and thus prevent flooding the sky with stars that we basically don't observe (in the numbers predicted by a pure GR theory). And it would eliminate the so-called "period spike" problem — the absence of a pile-up at minimum $P_o$ — as long as angular momentum loss is somewhat ideosyncratic, not strictly determined by $P_o$ (P01; Barker & Kolb 2003; King, Schenker, & Hameury 2003). Since mean accretion rates range by about an order of magnitude at essentially every $P_o$, the latter assumption has some plausibility. These are substantial returns for the investment of just one hypothesis!

### 7. SUMMARY

1. We report new orbital and superhump periods enabling a precise measurement of $\varepsilon$ in twenty dwarf novae. In some cases these reproduce or improve upon values previously published, or cited in advance of publication. We present all the power spectra and periodograms, so the reader can evaluate the detection's significance and discrimination of aliases.

2. The new $\varepsilon$s are generally unremarkable, showing a trend with $P_o$ consistent with the known correlation. Except at the extremities of $P_o$, $\varepsilon$ tends to run ~25% low compared to predictions assuming a ZAMS secondary. The observed scatter in $\varepsilon(P_o)$ is a measure of the scatter in the physical variables contained in the tidal theory of superhumps. We estimate a dispersion in $M_1$ not exceeding 24%, and a dispersion in $R_2$ (the secondary's radius at a given $P_o$) not exceeding 11%. So low a dispersion in $R_2$ indicates that the secondaries substantially follow a "main sequence", even though it is not quite consistent with the theoretical ZAMS.

3. It is likely that all these secondaries[6] are of essentially solar composition, since significant nuclear evolution would produce a large effect on $\varepsilon$ — including presumably a large scatter at a fixed $P_o$ — which is not seen.

4. For the most well-populated and well-behaved part of $\varepsilon(P_o)$ space, with $P_o$ between 0.06 and 0.085 d, we use the $\varepsilon(q)$ relation and the 46 data points to measure $R_2$. The method requires a value for $<M_1>$, which we take to be 0.75 $M_\odot$. CV secondaries are then found to be on average 18±6% larger than theoretical ZAMS stars at $t$=10 Gyr. This makes them substantially less massive, cooler, and fainter than predicted by theory.

5. The values of $R_2$ increasingly depart from the main sequence for shorter $P_o$, suggesting an origin in increasing departure from thermal equilibrium. The variance in $\varepsilon(P_o)$ at very short $P_o$ suggests that these departures from thermal equilibrium are highly ideosyncratic, resulting in a *range* of minimum periods — mostly in the range 75–85 minutes, but some perhaps as long as 3 hours.

Heavily implicated but unindicted co-conspirators in this enterprise include the many visual observers whose nightly patrols yield the timely announcements of freshly erupted dwarf novae. You know who you are out there. Rod Stubbings, Patrick Schmeer, Timo Kinnunen, and Gary Poyner are among those whose work defines the industry standards. This information is all sped along very effectively by the regular electronic announcements, alerts, and pleadings of the AAVSO and VSNET (from Kyoto University). We thank the NSF (AST00–98254 and AST99–87334 to J.P. and J.T.) for financial support.

---

[6] All the ones considered here. A few dwarf novae appear to have high helium abundance, which should drastically affect both $\varepsilon(P_o)$ and the range of $P_o$ accessible in evolution. These matters are discussed by Thorstensen et al. (2002).

# REFERENCES

Balayan, S.K. 1997, Ap., 40, 211.
Baraffe, I. 2002, in The Physics of Cataclysmic Variables and Related Objects, ed. B.T. Gaensicke, K. Beuermann, & K. Reinsch (San Francisco: Astronomical Society of the Pacific), p. 2.
Baraffe, I., Chabrier, G., Allard, F., & Hauschildt, P. 1998, A&A, 337, 403 (BCAH).
Barker, J. & Kolb, U. 2003, MNRAS, submitted.
Barwig, H., Mantel, K.-H., & Ritter, H. 1992, A&A, 266, L5.
Beuermann, K., Baraffe, I., Kolb, U., & Weichhold, M. 1998, A&A, 339, 518.
Bond, H.E., Kemper, E., & Mattei, J.A. 1982, ApJ, 260, L79.
Faulkner, J., Flannery, B.P., & Warner, B. 1972, ApJ, 175, L79.
Friend, M.T., Martin, J.S., Smith, R.C., & Jones, D.H.P. 1990, MNRAS, 246, 637.
Hazen, M.L. & Garnavich, P. 1999, JAAVSO, 27, 19.
Hellier, C. 2000, Cataclysmic Variable Stars: How and Why They Vary (Berlin: Springer), chap. 6.
Hirose, & Osaki, Y. 1990, PASJ, 42, 135.
Howell, S.B. & Szkody, P. 1988, PASP, 100, 224.
Ichikawa, S., Hirose, M., & Osaki, Y. 1993, PASJ, 45, 243.
Ishioka, R., Kato, T., Uemura, M., Iwamatsu, H., Matsumoto, K., Stubbings, R., Mennickent, R., Billings, G.W., Kiyota, S., Masi, G., Pietz, J., Novák, R., Martin, B.E., Oksanen, A., Moilanen, M., Torii, K., Kimugasa, K., & Kawakita, H. 2001, PASJ, 53, 905.
Jiang, X.J., Engels, D., Wei, J.Y., Tesch, F., & Hu, J.Y. 2000, A&A, 362, 263.
Kato, T. 1995, IBVS 4239.
Kato, T. 1996, IBVS 4369.
Kato, T. 2002, PASJ, 54, L11.
Kato, T. 2003, http://www.kusastro.kyoto-u.ac.jp/vsnet/DNe/cuvel0212.html.
Kato, T., Stubbings, R., Nelson, P., Santallo, R., Ishioka, R., Uemura, M., Sumi, T., Muraki, Y., Kilmartin, P., Bond, I., Noda, S., Yock, P., Hearnshaw, J.B., Monard, B., & Yamaoka, H. 2002, A&A, 395, 541.
Kato, T., Nogami, D., Masuda, S., & Baba, H. 1998, PASP, 110, 1400.
Kato, T., Sekine, Y., & Hirata, R. 2001, PASJ, 53, 1191.
Kato, T., Matsumoto, K., Nogami, D., Morikawa, K., & Kiyota, S. 2001, PASJ, 53, 893.
Kato, T., Bolt., G., Nelson, P., Monard, B., Stubbings, R., Pearce, A., Yamaoka, H., & Richards, T. 2003, MNRAS, in press.
King, A.R., Schenker, K., & Hameury, J.-M. 2003, MNRAS, submitted.
Kiyota, S. & Kato, T. 1998, IBVS 4644.
Koen, C. & O'Donoghue, D. 1992, ApSS, 101, 347.
Kolb, U. & Baraffe, I. 1999, MNRAS, 309, 1034.
Leibowitz, E.M., Mendelson, H., Bruch, A., Duerbeck, H.W., Seitter, W.C., & Richter, G.A. 1994, ApJ, 421, 771.
Liu, W. & Hu, J.Y. 2000, ApSS, 128, 387.
Lubow, S.H. 1991, ApJ, 381, 268.
Lubow, S.H. 1992, ApJ, 401, 317.
Markarian, B.E., & Stepanian, D.A. 1983, Afz, 19, 639.
Maza, J., Gonzalez, L.E., Wischniewsky, M., & Barrientos, F. 1992, PASP, 104, 1060.

Mennickent, R.E. & Diazm M. 1996, A&A, 309, 147.
Mennickent, R.E. & Tappert, H. 2001, A&A, 372, 563.
Mennickent, R.E., Matsumoto, K., & Arenas, J. 1999c, A&A, 348, 466.
Mennickent, R.E., Sterken, C., Gieren, W., & Unda, E. 1999b, A&A, 352, 239.
Mennickent, R.E., Patterson, J., O'Donoghue, D., Unda, E., Harvey, D., Vanmunster, T., & Bolt, G. 1999a, Ap&SS, 262, 1.
Misselt, K. & Shafter, A.W. 1995, AJ, 109, 1757.
Mukai, K., Mason, K.O., Howell, S.B., Allington-Smith, J., Callanan, P.J., Charles, P.A., Hassall, B.J.M., Naylor, T., Smale, A.P., & van Paradijs, J. 1990, MNRAS, 245, 385.
Munari, U. & Zwitter, T. 1998, A&AS, 128, 277.
Murray, J.R. 1998, MNRAS, 297, 323.
Murray, J.R. 2000, MNRAS, 314, L1.
Nogami, D., Baba, H., Katsura, M., & Kato, T. 2003, PASP, 55, 483.
Nogami, D., Kato, T., & Masuda, S. 1998, PASJ, 50, 411.
Nogami, D., Kato, T., Baba, H., & Masuda, S. 1998, PASJ, 50, L1.
Nogami, D., Engels, D., Gansicke, B.T., Pavlenko, E.P., Novák, R., & Reinsch, K. 2000, A&A, 364, 701.
Osaki, Y. 1996, PASP, 108, 39.
Osaki, Y. & Meyer, F. 2002, A&A, 383, 574.
Paczynski, B. 1981, AcA, 31, 1.
Patterson, J. 1981, ApJS, 45, 517.
Patterson, J. 1984, ApJS, 54, 443.
Patterson, J. 1998, PASP, 110, 1132 (P98).
Patterson, J. 2001, PASP, 113, 736 (P01).
Patterson, J., Bond, H.E., Grauer, A.D., Shafter, A.W., & Mattei, J.A. 1993, PASP, 105, 69
Patterson, J., Masi, G., Richmond, M.W., Martin, B., Beshore, E., Skillman, D.R., Kemp, J., Vanmunster, T., Rea, R., Allen, W., Davis, S., Davis, T., Henden, A.A., Starkey, D., Foote, J., Oksanen, A., Cook, L.M., Fried, R.E., Husar, D., Novák, R., Campbell, T., Robertson, J., Krajci, T., Pavlenko, E., Mirabal, N., Niarchos, P.G., Brettman, O., & Walker, S. 2002, PASP, 114, 721.
Renvoize, V., Baraffe, I., Kolb, U. & Ritter, H. 2002, A&A, 389, 485.
Schneider, D.P. & Young, P.J. 1980, ApJ, 238, 946
Semeniuk, I., Nalezyty, M., Gembara, P., & Kwast, T. 1997, AcA, 47, 299.
Skillman, D.R. & Patterson, J. 1993, ApJ, 417, 298.
Skillman, D.R., Krajci, T., Beshore, E., Patterson, J., Kemp, J., Starkey, D., Oksanen, A., Vanmunster, T., Martin, B., & Rea, R. 2002, PASP, 114, 630.
Smak, J. 1985, AcA, 35, 351.
Smith, D.A. & Dhillon, V.S. 1997, MNRAS, 301, 767.
Smith, R.C., Sarna, M.J., Catalán, M.S., & Jones, D.H.P. 1997, MNRAS, 287, 271.
Stolz, B. & Schoembs, R. 1984, A&A, 1332, 187.
Taylor, C.J., Thorstensen, J.R., & Patterson, J. 1999, PASP, 111, 184.
Thorstensen, J.R. & Fenton, W.H. 2002, PASP, 115, 37.
Thorstensen, J.R., Fenton, W.H., Patterson, J.O., Kemp, J., Krajci, T., & Baraffe, I. 2002, ApJ, 567, L49.
Thorstensen, J.R. & Freed, I. 1985, AJ, 90, 2082.
Thorstensen, J.R. & Taylor, C. 1997, PASP, 109, 1359.

Thorstensen, J.R., Patterson, J., Shambrook, A.A., & Thomas, G. 1996, PASP 108, 73.
Thorstensen, J.R., Taylor, C., & Kemp, J. 1998, PASP, 110, 1405.
Thorstensen, J.R., Wade, R.A., & Oke, J.B. 1986, 309, 721.
Uemura, M., Kato, T., Ishioka, R., Yamaoka, H., Schmeer, P., Krajci, T., Starkey, D.R., Torii, K., Kawai, N., Urata, Y., Kohama, M., Yoshida, A., Ayani, K., Kawabata, T., Tanabe, K., Matsumoto, K., Kiyota, S., Pietz, J., Vanmunster, T., Oksanen, A., & Giambersion, A. 2002, PASJ, 54, 599.
Vanmunster, T., Skillman, D.R., & Fried, R. 2000, IBVS 4940.
Vrielmann, S. & Offutt, W. 2003, MNRAS, 338, 165.
Vrielmann, S., Stiening, R.F., & Offutt, W. 2002, MNRAS, 334, 608.
Warner, B. 1995, Ap&SS, 226, 187.
Warner, B. & O'Donoghue, D. 1987, MNRAS, 224, 733.
Warner, B. & Woudt, P. 2001, MNRAS, 328, 159.
Webbink, R.F. 1990, in Accretion-Powered Compact Binaries, ed. C.W. Mauche (Cambridge: Cambridge University Press), p 177.
Whitehurst, R. 1988, MNRAS, 232, 35.
Wood, M.A., Montgomery, M.M., & Simpson, J.C. 2000, ApJ, 535, L39.

Table 1. Journal of Spectroscopy

| Date (UT) | N | HA start (hh:mm) | HA end (hh:mm) | Date (UT) | N | HA start (hh:mm) | HA end (hh:mm) | Date (UT) | N | HA start (hh:mm) | HA end (hh:mm) |
|---|---|---|---|---|---|---|---|---|---|---|---|
| *GX Cas:* | | | | *KV Dra:* | | | | *QW Ser:* | | | |
| 1996 Dec 28 | 5 | $+1:53$ | $+3:10$ | 2000 Apr 7 | 2 | $+0:14$ | $+0:24$ | 2001 Mar 24 | 2 | $-2:40$ | $-2:28$ |
| 1996 Dec 29 | 15 | $+1:05$ | $+2:48$ | 2000 Apr 8 | 15 | $-5:36$ | $+2:47$ | 2001 Mar 25 | 7 | $-2:43$ | $+1:05$ |
| 1996 Dec 30 | 13 | $+0:13$ | $+4:49$ | 2000 Apr 9 | 4 | $+1:19$ | $+1:49$ | 2001 Mar 26 | 4 | $-0:15$ | $+1:35$ |
| 1997 Jan 2 | 1 | $+1:07$ | $\cdots$ | 2000 Apr 10 | 4 | $-5:34$ | $+2:37$ | 2001 Mar 27 | 8 | $-3:33$ | $+1:50$ |
| 1997 Jan 3 | 7 | $+1:46$ | $+2:34$ | 2000 Apr 11 | 6 | $-4:16$ | $-3:31$ | 2001 Mar 28 | 2 | $-0:02$ | $+0:10$ |
| 1997 Sep 20 | 1 | $+1:48$ | $\cdots$ | 2000 Jul 4 | 19 | $+0:57$ | $+5:56$ | *BC UMa:* | | | |
| *TU Crt:* | | | | 2000 Jul 5 | 10 | $+2:00$ | $+3:56$ | 1999 Jun 4 | 1 | $+1:56$ | $\cdots$ |
| 2000 Jan 6 | 11 | $+0:04$ | $+1:27$ | *RZ Leo:* | | | | 1999 Jun 9 | 18 | $+1:52$ | $+5:49$ |
| 2000 Jan 7 | 13 | $-3:02$ | $+1:19$ | 1998 Mar 21 | 24 | $-1:07$ | $+3:44$ | 1999 Jun 10 | 5 | $+5:07$ | $+5:51$ |
| 2000 Jan 8 | 29 | $-3:12$ | $+2:05$ | 1998 Mar 22 | 17 | $-2:16$ | $+1:08$ | 1999 Jun 11 | 5 | $+3:59$ | $+4:35$ |
| 2000 Jan 9 | 4 | $-2:52$ | $-2:31$ | 1998 Mar 23 | 4 | $+2:30$ | $+3:31$ | 1999 Jun 13 | 12 | $+2:06$ | $+3:48$ |
| 2000 Jan 10 | 24 | $-0:35$ | $+2:14$ | 1998 Mar 24 | 7 | $+2:04$ | $+3:23$ | *KS UMa:* | | | |
| *HO Del:* | | | | 1998 Mar 25 | 4 | $-3:34$ | $-2:12$ | 1999 Jun 2 | 29 | $+2:14$ | $+5:25$ |
| 1996 Sep 28 | 4 | $+0:58$ | $+2:20$ | *RZ Sge:* | | | | 1999 Jun 3 | 21 | $+2:45$ | $+4:51$ |
| 1996 Sep 29 | 12 | $+0:38$ | $+4:14$ | 1998 Sep 10 | 20 | $+1:20$ | $+4:26$ | 1999 Jun 4 | 7 | $+2:27$ | $+3:02$ |
| 1996 Sep 30 | 29 | $-0:22$ | $+4:50$ | 1999 Jun 2 | 2 | $-3:54$ | $-3:46$ | *HS Vir:* | | | |
| 1996 Oct 1 | 14 | $-0:49$ | $+1:48$ | 1999 Jun 3 | 36 | $-4:14$ | $+0:41$ | 1996 Apr 6 | 1 | $-0:42$ | $\cdots$ |
| 1996 Oct 2 | 12 | $+0:28$ | $+2:10$ | 1999 Jun 4 | 12 | $-3:31$ | $+0:14$ | 1996 Apr 7 | 20 | $-0:40$ | $+3:33$ |
| 1997 Sep 21 | 1 | $+0:39$ | $\cdots$ | 1999 Jun 10 | 2 | $-0:20$ | $-0:13$ | 1996 Apr 8 | 9 | $-2:33$ | $+3:38$ |
| | | | | 1999 Jun 11 | 2 | $-0:20$ | $-0:13$ | 1996 Apr 9 | 9 | $-3:27$ | $+3:18$ |
| | | | | 1999 Jun 13 | 7 | $-0:55$ | $-0:14$ | | | | |
| | | | | 1999 Jun 14 | 10 | $+0:03$ | $+1:07$ | | | | |

Table 2. Emission Features

| Feature | E.W.[a] (Å) | Flux[b] ($10^{-15}$ erg cm$^{-2}$ s$^1$) | FWHM[c] (Å) |
|---|---|---|---|
| GX Cas: | | | |
| H$\beta$ | 38 | 54 | 36 |
| HeI $\lambda$5015 | 5 | 8 | 53 |
| Fe $\lambda$5169 | 8 | 12 | 45 |
| HeI $\lambda$5876 | 11 | 14 | 47 |
| H$\alpha$ | 87 | 97 | 40 |
| TU Crt: | | | |
| H$\beta$ | 71 | 104 | 33 |
| HeI $\lambda$4921 | 7 | 10 | 33 |
| HeI $\lambda$5015 | 8 | 12 | 35 |
| Fe $\lambda$5169 | 9 | 14 | 36 |
| Fe $\lambda$5169 | 9 | 14 | 36 |
| Fe $\lambda$5169 | 9 | 14 | 38 |
| HeI $\lambda$5876 | 28 | 36 | 46 |

– 24 –

Table 2—Continued

| Feature | E.W.[a] (Å) | Flux[b] ($10^{-15}$ erg cm$^{-2}$ s$^1$) | FWHM [c] (Å) |
|---|---|---|---|
| H$\alpha$ | 145 | 172 | 39 |
| HeI $\lambda$6678 | 14 | 16 | 60 |
| HO Del: | | | |
| H$\beta$ | 74 | 75 | 24 |
| HeI $\lambda$5015 | 10 | 9 | 30 |
| Fe $\lambda$5169 | 6 | 7 | 18 |
| HeI $\lambda$5876 | 34 | 29 | 29 |
| H$\alpha$ | 134 | 110 | 24 |
| HeI $\lambda$6678 | 12 | 10 | 30 |
| HeI $\lambda$7067 | 14 | 14 | 43 |
| KV Dra (2000 July): | | | |
| H$\gamma$ | 46 | 337 | 36 |
| HeI $\lambda$4471 | 15 | 92 | 34 |
| H$\beta$ | 58 | 274 | 33 |

Table 2—Continued

| Feature | E.W.[a] (Å) | Flux[b] ($10^{-15}$ erg cm$^{-2}$ s$^{1}$) | FWHM [c] (Å) |
|---|---|---|---|
| HeI $\lambda$5015 | 7 | 28 | 34 |
| Fe $\lambda$5169 | 5 | 19 | 33 |
| HeI $\lambda$5876 | 25 | 68 | 40 |
| H$\alpha$ | 91 | 230 | 38 |
| HeI $\lambda$6678 | 12 | 29 | 48 |
| HeI $\lambda$7067 | 10 | 22 | 51 |
| RZ Leo: | | | |
| H$\beta$ | 36 | 44 | 37 |
| HeI $\lambda$5876 | 12 | 10 | 59 |
| H$\alpha$ | 102 | 73 | 47 |
| RZ Sge: | | | |
| H$\gamma$ | 33 | 119 | 30 |
| HeI $\lambda$4471 | 4 | 16 | 23 |
| H$\beta$ | 42 | 132 | 29 |

Table 2—Continued

| Feature | E.W.[a] (Å) | Flux[b] ($10^{-15}$ erg cm$^{-2}$ s$^1$) | FWHM [c] (Å) |
|---|---|---|---|
| HeI λ5015 | 4 | 14 | 36 |
| HeI λ5876 | 17 | 46 | 40 |
| Hα | 85 | 207 | 36 |
| HeI λ6678 | 9 | 22 | 48 |
| HeI λ7067 | 5 | 11 | 41 |
| QW Ser: | | | |
| Hγ | 15 | 51 | 37 |
| Hβ | 27 | 73 | 43 |
| Fe λ5169 | 6 | 15 | 37 |
| HeI λ5876 | 10 | 22 | 58 |
| Hα | 62 | 108 | 48 |
| HeI λ6678 | 5 : | 8 : | 89: |

[a]Emission equivalent widths are counted as positive.

[b]Absolute line fluxes are uncertain by a factor of about 2, but relative fluxes of strong lines are estimated accurate to $\sim 10$ per cent.

[c]From Gaussian fits.

Table 3.  Hα Radial Velocities

| Time (HJD 24...) | v km s$^{-1}$ | Time (HJD 24...) | v km s$^{-1}$ | Time (HJD 24...) | v km s$^{-1}$ | Time (HJD 24...) | v km s$^{-1}$ |
|---|---|---|---|---|---|---|---|
| GX Cas: two gaussian, separation 50 Å, width 12 Å ||||||||
| 50711.9237 | −251 | 50446.6418 | 80 | 50447.5853 | −95 | 50447.7720 | −170 |
| 50445.6558 | 50 | 50446.6462 | 54 | 50447.5897 | −22 | 50450.6099 | 9 |
| 50445.6615 | 101 | 50446.6522 | −110 | 50447.5940 | 71 | 50451.6342 | −2 |
| 50445.6722 | 67 | 50446.6566 | 80 | 50447.7215 | 23 | 50451.6392 | 21 |
| 50445.7023 | −98 | 50446.6610 | 21 | 50447.7259 | −93 | 50451.6443 | 15 |
| 50445.7088 | −236 | 50446.6654 | −84 | 50447.7303 | 46 | 50451.6494 | −46 |
| 50446.6193 | −48 | 50446.6698 | −88 | 50447.7434 | 41 | 50451.6545 | −110 |
| 50446.6237 | 0 | 50446.6741 | −144 | 50447.7517 | −147 | 50451.6620 | 22 |
| 50446.6286 | −5 | 50446.6865 | −58 | 50447.7567 | −168 | 50451.6671 | −93 |
| 50446.6330 | 118 | 50446.6909 | −139 | 50447.7618 | −112 | | |
| 50446.6374 | 67 | 50447.5809 | −194 | 50447.7669 | −109 | | |
| TU Crt: two gaussian, separation = 44 Å, width = 16 Å ||||||||
| 51549.9826 | 14 | 51551.0225 | −37 | 51552.0068 | −19 | 51553.9760 | −22 |
| 51549.9874 | 22 | 51551.0272 | −76 | 51552.0115 | −79 | 51553.9808 | 13 |
| 51549.9921 | 64 | 51551.0320 | −29 | 51552.0163 | −47 | 51553.9855 | −53 |
| 51549.9969 | 89 | 51551.8419 | 23 | 51552.0210 | −93 | 51553.9903 | −69 |
| 51550.0016 | 77 | 51551.8467 | −2 | 51552.0257 | −98 | 51553.9950 | −65 |
| 51550.0063 | 74 | 51551.8515 | −18 | 51552.0378 | −25 | 51553.9998 | −60 |
| 51550.0212 | 136 | 51551.8562 | −5 | 51552.0425 | 55 | 51554.0076 | −34 |

Table 3—Continued

| Time (HJD 24…) | $v$ km s$^{-1}$ | Time (HJD 24…) | $v$ km s$^{-1}$ | Time (HJD 24…) | $v$ km s$^{-1}$ | Time (HJD 24…) | $v$ km s$^{-1}$ |
|---|---|---|---|---|---|---|---|
| 51550.0259 | 76 | 51551.8609 | −25 | 51552.0473 | 106 | 51554.0124 | 12 |
| 51550.0307 | 63 | 51551.8656 | −41 | 51552.0520 | 71 | 51554.0172 | 62 |
| 51550.0354 | 25 | 51551.8733 | −4 | 51552.0568 | 93 | 51554.0219 | 50 |
| 51550.0402 | 15 | 51551.8780 | 38 | 51552.0615 | 130 | 51554.0266 | 77 |
| 51550.8511 | 128 | 51551.8828 | 82 | 51552.8532 | −26 | 51554.0314 | 104 |
| 51550.8573 | 1 | 51551.8875 | 76 | 51552.8579 | −100 | 51554.0382 | 75 |
| 51550.8634 | −48 | 51551.8922 | 93 | 51552.8627 | 40 | 51554.0430 | 61 |
| 51550.8695 | 3 | 51551.8970 | 129 | 51552.8674 | 94 | 51554.0477 | 110 |
| 51550.8757 | −44 | 51551.9039 | 104 | 51553.9454 | 119 | 51554.0525 | 38 |
| 51550.8857 | −31 | 51551.9086 | 93 | 51553.9502 | 85 | 51554.0572 | 32 |
| 51550.8918 | 31 | 51551.9134 | 103 | 51553.9549 | 115 | 51554.0620 | 14 |
| 51550.8979 | 77 | 51551.9181 | 63 | 51553.9597 | 70 | | |
| 51551.0130 | 48 | 51551.9228 | 2 | 51553.9644 | 64 | | |
| 51551.0177 | 60 | 51551.9276 | −16 | 51553.9691 | 23 | | |

| HO Del: two gaussian, separation = 70 Å, width = 18 Å ||||||||
|---|---|---|---|---|---|---|---|
| 50354.6926 | 12 | 50356.6495 | 1 | 50356.7833 | −136 | 50357.6742 | 40 |
| 50354.6986 | 111 | 50356.6555 | −74 | 50356.7982 | −46 | 50357.6803 | −105 |
| 50354.7047 | −57 | 50356.6616 | −46 | 50356.8075 | −31 | 50357.7004 | −2 |
| 50354.7493 | 78 | 50356.6735 | −12 | 50356.8135 | −15 | 50357.7065 | 5 |
| 50355.6758 | −20 | 50356.6796 | −67 | 50356.8196 | 19 | 50357.7185 | −79 |
| 50355.6880 | 88 | 50356.6856 | 4 | 50356.8292 | −77 | 50358.6601 | −34 |
| 50355.7205 | −204 | 50356.6917 | 40 | 50356.8352 | −8 | 50358.6662 | −10 |

Table 3—Continued

| Time (HJD 24...) | $v$ km s$^{-1}$ | Time (HJD 24...) | $v$ km s$^{-1}$ | Time (HJD 24...) | $v$ km s$^{-1}$ | Time (HJD 24...) | $v$ km s$^{-1}$ |
|---|---|---|---|---|---|---|---|
| 50355.7265 | −136 | 50356.6977 | −20 | 50356.8413 | 3 | 50358.6723 | −120 |
| 50355.7675 | −90 | 50356.7140 | −31 | 50356.8473 | −68 | 50358.6783 | −32 |
| 50355.7735 | −4 | 50356.7200 | −156 | 50357.6099 | −12 | 50358.6844 | 15 |
| 50355.7948 | −167 | 50356.7261 | −32 | 50357.6160 | 13 | 50358.6904 | 36 |
| 50355.8009 | 7 | 50356.7322 | −64 | 50357.6220 | −115 | 50358.7009 | −17 |
| 50355.8069 | −33 | 50356.7382 | 38 | 50357.6341 | −13 | 50358.7069 | −59 |
| 50355.8130 | 96 | 50356.7443 | −74 | 50357.6402 | 112 | 50358.7130 | 29 |
| 50355.8190 | 58 | 50356.7591 | 80 | 50357.6500 | −120 | 50358.7190 | 48 |
| 50355.8251 | 0 | 50356.7652 | 2 | 50357.6560 | −70 | 50358.7251 | −41 |
| 50356.6313 | 135 | 50356.7712 | −55 | 50357.6621 | −101 | 50358.7311 | −105 |
| 50356.6373 | −52 | 50356.7773 | −131 | 50357.6681 | −83 | 50712.6992 | −217 |
| KV Dra: two gaussian, separation = 50 Å, width = 16 Å | | | | | | | |
| 51641.8956 | −29 | 51642.9929 | −84 | 51645.7288 | 16 | 51729.8633 | −37 |
| 51641.9028 | 8 | 51642.9990 | −113 | 51729.6825 | −55 | 51729.8694 | −53 |
| 51642.6505 | −131 | 51643.9353 | −63 | 51729.6887 | −29 | 51729.8755 | −103 |
| 51642.6570 | −68 | 51643.9435 | −137 | 51729.6948 | −123 | 51729.8836 | −52 |
| 51642.8763 | −90 | 51643.9496 | −42 | 51729.7009 | −135 | 51729.8895 | −82 |
| 51642.8825 | −143 | 51643.9557 | −15 | 51729.7162 | 9 | 51730.7237 | −8 |
| 51642.8887 | −63 | 51644.6460 | −17 | 51729.7479 | 9 | 51730.7298 | 9 |
| 51642.8948 | −69 | 51644.6543 | −24 | 51729.7540 | −38 | 51730.7359 | −9 |
| 51642.9009 | 1 | 51644.6604 | −30 | 51729.7601 | −84 | 51730.7420 | −23 |
| 51642.9094 | 28 | 51644.9862 | −96 | 51729.7662 | −51 | 51730.7731 | −12 |

Table 3—Continued

| Time (HJD 24...) | $v$ km s$^{-1}$ | Time (HJD 24...) | $v$ km s$^{-1}$ | Time (HJD 24...) | $v$ km s$^{-1}$ | Time (HJD 24...) | $v$ km s$^{-1}$ |
|---|---|---|---|---|---|---|---|
| 51642.9216 | −11 | 51645.6979 | −106 | 51729.7723 | 8 | 51730.7792 | 36 |
| 51642.9278 | −91 | 51645.7044 | −118 | 51729.7784 | 5 | 51730.7853 | −24 |
| 51642.9339 | −123 | 51645.7105 | −62 | 51729.8449 | 19 | 51730.7915 | 23 |
| 51642.9807 | 20 | 51645.7166 | −53 | 51729.8511 | −12 | 51730.7976 | −4 |
| 51642.9868 | −150 | 51645.7227 | −21 | 51729.8572 | 19 | 51730.8037 | −50 |
| RZ Leo : two gaussian, separation = 32 Å, width = 12 Å | | | | | | | |
| 50893.7524 | −331 | 50893.8855 | −214 | 50894.7607 | −79 | 50896.8926 | −122 |
| 50893.7578 | −201 | 50893.8909 | −162 | 50894.7662 | −97 | 50896.8980 | 38 |
| 50893.7688 | −3 | 50893.8964 | −377 | 50894.7716 | −95 | 50896.9142 | 120 |
| 50893.7742 | 4 | 50893.9042 | −456 | 50894.7840 | 80 | 50896.9251 | −131 |
| 50893.7797 | −32 | 50893.9096 | 73 | 50894.8113 | −428 | 50896.9305 | −144 |
| 50893.7909 | 106 | 50893.9151 | 27 | 50894.8213 | −192 | 50896.9360 | −392 |
| 50893.7964 | −220 | 50893.9260 | −42 | 50894.8268 | 96 | 50897.6445 | 54 |
| 50893.8073 | −493 | 50893.9427 | 112 | 50894.8323 | −10 | 50897.6554 | −169 |
| 50893.8127 | −281 | 50893.9537 | −160 | 50894.8377 | −119 | 50897.6609 | −5 |
| 50893.8237 | −473 | 50893.9591 | −214 | 50894.8432 | −12 | 50897.6663 | −51 |
| 50893.8312 | −223 | 50894.7069 | 109 | 50894.8486 | −68 | 50897.6854 | −101 |
| 50893.8366 | 109 | 50894.7178 | −271 | 50895.9022 | −101 | 50897.6909 | −122 |
| 50893.8421 | 163 | 50894.7233 | −207 | 50895.9076 | −254 | 50897.6964 | −361 |
| 50893.8476 | −45 | 50894.7287 | −242 | 50895.9131 | −90 | 50897.7018 | −337 |
| 50893.8530 | −96 | 50894.7342 | −234 | 50895.9444 | −166 | | |
| 50893.8585 | −196 | 50894.7498 | 78 | 50896.8817 | 175 | | |

Table 3—Continued

| Time (HJD 24...) | $v$ km s$^{-1}$ | Time (HJD 24...) | $v$ km s$^{-1}$ | Time (HJD 24...) | $v$ km s$^{-1}$ | Time (HJD 24...) | $v$ km s$^{-1}$ |
|---|---|---|---|---|---|---|---|
| 50893.8800 | −256 | 50894.7553 | 135 | 50896.8871 | −59 | | |
| QW Ser: two gaussian, separation = 52 Å, width = 14 Å | | | | | | | |
| 51992.8405 | −72 | 51993.9736 | −84 | 51995.0122 | −74 | 51995.8357 | −89 |
| 51992.8494 | −65 | 51993.9825 | −20 | 51995.7959 | 118 | 51995.8419 | −89 |
| 51993.8361 | −35 | 51993.9938 | 46 | 51995.8045 | 74 | 51996.0199 | 96 |
| 51993.8454 | 55 | 51994.9358 | −70 | 51995.8118 | −1 | 51996.9398 | −30 |
| 51993.8540 | 60 | 51994.9449 | −95 | 51995.8179 | 18 | 51996.9479 | −48 |
| 51993.9630 | 10 | 51994.9994 | −14 | 51995.8267 | −81 | | |
| RZ Sge: two gaussian, separation = 52 Å, width = 16 Å | | | | | | | |
| 51066.7354 | −23 | 51066.8640 | 24 | 51332.9736 | −43 | 51343.9273 | −66 |
| 51066.7402 | 3 | 51332.7933 | −92 | 51332.9785 | −62 | 51343.9320 | −40 |
| 51066.7449 | −23 | 51332.8478 | −139 | 51333.9378 | −104 | 51343.9367 | −11 |
| 51066.7497 | −122 | 51332.8526 | −89 | 51333.9426 | −111 | 51343.9415 | 14 |
| 51066.7545 | −58 | 51332.8574 | −127 | 51333.9474 | −75 | 51343.9462 | 18 |
| 51066.7592 | −93 | 51332.8622 | −96 | 51333.9522 | −71 | 51343.9510 | −14 |
| 51066.7640 | −94 | 51332.8748 | 33 | 51333.9570 | −90 | 51343.9557 | 14 |
| 51066.7708 | −137 | 51332.8843 | −12 | 51339.9172 | 22 | 51343.9624 | 6 |
| 51066.7755 | −45 | 51332.8891 | −27 | 51339.9220 | 20 | 51343.9672 | −11 |
| 51066.7803 | −43 | 51332.9239 | −98 | 51340.9147 | −96 | 51333.8013 | −91 |
| 51066.7851 | −37 | 51332.9287 | −95 | 51340.9194 | −70 | 51333.8061 | −139 |

Table 3—Continued

| Time (HJD 24…) | $v$ km s$^{-1}$ | Time (HJD 24…) | $v$ km s$^{-1}$ | Time (HJD 24…) | $v$ km s$^{-1}$ | Time (HJD 24…) | $v$ km s$^{-1}$ |
|---|---|---|---|---|---|---|---|
| 51066.7899 | −11 | 51332.9334 | −47 | 51342.8851 | −71 | 51333.8109 | −104 |
| 51066.7946 | 78 | 51332.9382 | −93 | 51342.8898 | −101 | 51333.8157 | −121 |
| 51066.7994 | 12 | 51332.9430 | −10 | 51342.8946 | −113 | 51333.8204 | −61 |
| 51066.8354 | −113 | 51332.9478 | −37 | 51342.8993 | −116 | 51333.8252 | −7 |
| 51066.8402 | −130 | 51332.9545 | 8 | 51342.9041 | −48 | 51333.8300 | 5 |
| 51066.8450 | −44 | 51332.9593 | 37 | 51342.9088 | −56 | | |
| 51066.8497 | −73 | 51332.9641 | 14 | 51342.9136 | 2 | | |
| 51066.8545 | −99 | 51332.9688 | −17 | 51343.9225 | −78 | | |

| BC UMa: two gaussian, separation = 52 Å, width = 14 Å ||||||||
|---|---|---|---|---|---|---|---|
| 51333.6839 | −28 | 51338.7626 | −50 | 51339.8242 | 12 | 51342.6906 | 23 |
| 51338.6674 | 12 | 51338.7687 | −79 | 51339.8303 | −37 | 51342.6967 | 12 |
| 51338.6722 | −40 | 51338.8007 | 28 | 51340.7499 | 32 | 51342.7054 | −31 |
| 51338.6776 | 18 | 51338.8068 | 5 | 51340.7561 | 26 | 51342.7115 | −80 |
| 51338.6839 | 27 | 51338.8130 | 5 | 51340.7622 | −45 | 51342.7177 | −96 |
| 51338.6900 | −47 | 51338.8191 | 6 | 51340.7683 | −37 | 51342.7238 | −74 |
| 51338.6961 | −39 | 51338.8252 | −64 | 51340.7745 | −94 | 51342.7300 | −70 |
| 51338.7380 | −13 | 51338.8314 | −115 | 51342.6660 | −20 | 51342.7361 | −2 |
| 51338.7442 | 58 | 51339.7996 | 8 | 51342.6722 | −33 | | |
| 51338.7503 | 13 | 51339.8119 | 9 | 51342.6783 | 17 | | |
| 51338.7565 | −34 | 51339.8180 | 3 | 51342.6845 | 18 | | |

KS UMa : two gaussian, separation = 32 Å, width = 10 Å

Table 3—Continued

| Time (HJD 24...) | $v$ km s$^{-1}$ | Time (HJD 24...) | $v$ km s$^{-1}$ | Time (HJD 24...) | $v$ km s$^{-1}$ | Time (HJD 24...) | $v$ km s$^{-1}$ |
|---|---|---|---|---|---|---|---|
| 51331.6371 | −58 | 51331.7112 | −49 | 51332.6597 | −45 | 51332.7265 | −16 |
| 51331.6434 | −48 | 51331.7152 | −3 | 51332.6637 | −4 | 51332.7306 | −22 |
| 51331.6474 | −35 | 51331.7193 | −25 | 51332.6678 | 1 | 51332.7347 | −17 |
| 51331.6515 | −11 | 51331.7234 | −30 | 51332.6719 | 25 | 51332.7388 | 5 |
| 51331.6556 | −4 | 51331.7283 | 7 | 51332.6760 | 55 | 51332.7429 | 23 |
| 51331.6597 | 15 | 51331.7324 | 19 | 51332.6801 | 48 | 51333.6402 | 59 |
| 51331.6638 | 44 | 51331.7365 | 61 | 51332.6873 | 36 | 51333.6443 | 70 |
| 51331.6679 | 47 | 51331.7438 | 45 | 51332.6914 | 41 | 51333.6484 | 53 |
| 51331.6781 | 31 | 51331.7487 | 35 | 51332.6955 | 46 | 51333.6525 | 47 |
| 51331.6822 | 40 | 51331.7527 | 16 | 51332.6996 | 14 | 51333.6566 | 16 |
| 51331.6862 | −4 | 51331.7568 | −12 | 51332.7037 | 16 | 51333.6607 | 22 |
| 51331.6903 | 11 | 51331.7609 | −24 | 51332.7077 | 2 | 51333.6647 | −17 |
| 51331.6944 | −24 | 51331.7650 | −30 | 51332.7118 | −28 | | |
| 51331.6985 | −41 | 51331.7691 | −49 | 51332.7184 | −29 | | |
| 51331.7051 | −65 | 51332.6556 | −60 | 51332.7225 | −41 | | |

HS Vir: two gaussian, separation = 60 Å, width = 18 Å

| 50179.8174 | −34 | 50180.9221 | −14 | 50180.9910 | −75 | 50182.6950 | −16 |
| 50180.8159 | 4 | 50180.9334 | 0 | 50181.7348 | 37 | 50182.7011 | 1 |
| 50180.8198 | −134 | 50180.9395 | 6 | 50181.7416 | 63 | 50182.7072 | 8 |
| 50180.8297 | −42 | 50180.9455 | −5 | 50181.7477 | −44 | 50182.7133 | −31 |
| 50180.8358 | −102 | 50180.9516 | 46 | 50181.7537 | −18 | 50182.7499 | −82 |

Table 3—Continued

| Time (HJD 24...) | $v$ km s$^{-1}$ | Time (HJD 24...) | $v$ km s$^{-1}$ | Time (HJD 24...) | $v$ km s$^{-1}$ | Time (HJD 24...) | $v$ km s$^{-1}$ |
|---|---|---|---|---|---|---|---|
| 50180.8418 | $-59$ | 50180.9576 | 85 | 50181.9525 | 21 | 50182.7560 | $-110$ |
| 50180.8979 | $-9$ | 50180.9668 | 33 | 50181.9587 | 78 | 50182.9629 | 45 |
| 50180.9039 | $-51$ | 50180.9729 | $-1$ | 50181.9799 | $-74$ | 50182.9690 | $-19$ |
| 50180.9100 | $-76$ | 50180.9789 | $-20$ | 50181.9860 | $-133$ | 50182.9751 | $-18$ |
| 50180.9160 | $-66$ | 50180.9850 | $-47$ | 50181.9921 | $-49$ | | |

Table 4. Fits to Radial Velocities

| Data set | $T_0$[a] | $P$ (d) | $K$ (km s$^{-1}$) | $\gamma$ (km s$^{-1}$) | $\sigma$ (km s$^{-1}$) | $N$ |
|---|---|---|---|---|---|---|
| GX Cas | 50447.776(3) | 0.08902(16) | 96(18) | −31(13) | 41 | 61 |
| TU Crt | 51552.0430(12) | 0.08209(9) | 85(9) | 30(6) | 81 | 27 |
| HO Del | 50356.867(3) | 0.06266(16) | 54(15) | −26(11) | 71 | 56 |
| KV Dra | 51645.7149(16) | 0.0587418(14):: | 54(9) | −36(6) | 60 | 30 |
| RZ Leo | 50894.824(4) | 0.0761(2) | 129(35) | −132(28) | 65 | 129 |
| RZ Sge | 51332.8021(10) | 0.0682803(5):: | 62(5) | −43(4) | 73 | 23 |
| QW Ser | 51994.9596(14) | 0.07453(10) | 85(8) | 10(6) | 23 | 25 |
| BC UMa | 51339.8549(11) | 0.06261(4) | 56(6) | −29(4) | 41 | 20 |
| KS UMa | 51332.6724(11) | 0.06796(10) | 47(5) | 8(3) | 57 | 16 |
| HS Vir | 50181.704(3) | 0.0769(2) | 59(12) | −12(8) | 39 | 32 |

[a]Blue-to-red crossing, HJD −2450000.

TABLE 5
SINUSOIDAL FITS TO S-WAVES

| Star | Phase[a] | Amplitude (km/s) |
|---|---|---|
| GX Cas | 0.341(14) | 542(47) |
| TU Crt | 0.296(17) | 629(68) |
| RZ Leo | 0.286(17) | 775(89) |
| BC UMa | 0.203(6) | 744(29) |

[a]Phase relative to the epochs listed in Table 4.

TABLE 6
CBA PHOTOMETRY CAMPAIGNS

| Star | $V_{max}$ | $V_{min}$ | $T_{rec}$ (d)[a] | Nights/hours | Telescopes[b] |
|---|---|---|---|---|---|
| BC UMa | 11.8 | 18.4 | 1000 | 31/170 | 3,9,13,2,5,7,6 |
| KV Dra | 13.2 | >16 | (400) | 18/94 | 3,5,1,13 |
| KS UMa | 12.5 | 16.3 | 260 | 18/95 | 5,1,3,13 |
| RZ Leo | 12 | 19 | (2000) | 16/60 | 5,7,1,10,4 |
| HO Del | 13.8 | 19 | (1000) | 6/24 | 5,4,1 |
| HS Vir | 13.4 | 16.2 | 370 | 11/55 | 1,2,11 |
| GX Cas | 13 | 17.5 | 360 | 14/88 | 1,2 |
| QW Ser | 12.6 | 17 | 300 | 6/23 | 1,10,13 |
| RZ Sge | 12.3 | 17.7 | 270 | 8/35 | 11,5,3 |
| LL And | 13.5 | 20.0 | >2000 | 6/22 | 13 |
| KV And | 14 | 19.5 | 240 | 10/37 | 1,2,9,13 |
| WX Cet | 11.8 | 18.3 | 900 | 21/110 | 11,4,3,14 |
| MM Hya | 13.2 | 16.2 | 400 | 24/94 | 13,2,4,5,16 |
| AO Oct | 14.2 | 20.2 |  | 15/75 | 15 |
| V2051 Oph | 13 | 15.3 | 450 | 26/86 | 17,8,15,2,19,18 |
| NY Ser | 14.7 | 18.5 | 60–100 | 61/287 | 13,1,2,3,6 |
| KK Tel | 13.7 |  | ((600)) | 14/75 | 12,15 |
| HV Vir | 12 | 19 | 3000 | 42/170 | 22,6,23,24,15,5 |
| RX J1155–56 | 11.5 | 14.9 |  | 34/216 | 19,20,18,17,21 |

[a]$T_{rec}$ is the estimated recurrence time between superoutbursts. Parentheses denote extra uncertainty.

[b]Telescopes: 1 = CBA–East 66 cm, D Skillman; 2 = CBA–Tucson 35 cm, D. Harvey; 3 = CBA–Flagstaff 41 cm, R. Fried; 4 = CBA–Denmark 25 cm, L. Jensen; 5 = CBA–Belgium 35 cm, T. Vanmunster; 6 = CBA–Concord 46 cm, L. Cook; 7 = KUC 32 cm, B. Martin; 8 = CBA–Tamworth 46 cm, G. Garradd; 9 = 60 cm, E. Pavlenko; 10 = CBA–Italy 25 cm, G. Masi; 10 = NCO 41 cm, R. Novák; 11 = CTIO 1 m, J. Patterson; 12 = CBA–Pakuranga 35 cm, J.McCormick & F.Velthius; 13 = MDM 1.3 m, J.Kemp; 14 = CBA–Illinois 20 cm, J. Gunn; 15 = CTIO 0.9 m, J. Kemp; 16 = SAAO 75 cm, D. O'Donoghue; 17 = CBA–Awanui 25 cm, S. Walker; 18 = CBA–Perth 35 cm, G. Bolt; 19 = CBA–Pretoria 35 cm, B. Monard; 20 = CBA–Nelson 35 cm, R. Rea; 21=CBA–Townsville 20 cm, N. Butterworth; 22 = CBA–New Mexico 28 cm, T. Krajci; 23=CBA–Utah 51 cm, J. Foote; 24 = CBA–Colorado 35 cm, E. Beshore.

TABLE 7
PERIOD SUMMARY

| Star | $P_{orb}$ (d) | $P_{sh}$ (d) | $\varepsilon$ | References[a] |
|---|---|---|---|---|
| GX Cas | 0.08902(16) | 0.09302(17) | 0.0449(25) | 1, 2 |
| TU Crt | 0.08209(9) | 0.08535(18) | 0.0397(22) | 1, 3 |
| HO Del | 0.06266(16) | 0.06439(18) | 0.0276(35) | 1 |
| KV Dra | 0.05876(7) | 0.06013(11) | 0.0233(22) | 1, 5 |
| RZ Leo | 0.0760383(4) | 0.07868(19) | 0.0347(25) | 1, 6, 7 |
| QW Ser | 0.07453(10) | 0.0770(3) | 0.0331(40) | 1 |
| RZ Sge | 0.068282(18) | 0.07037(19) | 0.0306(28) | 1, 8, 9, 10 |
| BC UMa | 0.062605(11) | 0.06452(9) | 0.0306(14) | 1 |
| KS UMa | 0.06796(10) | 0.0696(2) | 0.0241(30) | 1 |
| HS Vir | 0.0769(2) | 0.08045(19) | 0.0462(35) | 1, 12, 13 |
| LL And | 0.055053(5) | 0.0567 | 0.030 | 1, 14 |
| KV And | 0.07326(21) | 0.07435(12) | 0.0145(32) | 1, 15 |
| WX Cet | 0.05829(4) | 0.05945(7) | 0.0199(15) | 1, 16, 17 |
| MM Hya | 0.057590(2) | 0.05868(7) | 0.0189(14) | 1 |
| AO Oct | 0.06557(13) | 0.06716(14) | 0.0242(39) | 1 |
| V2051 Oph | 0.062427(<1) | 0.06418(16) | 0.0281(25) | 1, 18, 19 |
| NY Ser | 0.09775(19) | 0.10384(32) | 0.0623(35) | 1, 20 |
| KK Tel | 0.08453(21) | 0.08764(15) | 0.0368(31) | 1, 21 |
| RX J1155–56 | 0.06028(10) | 0.06215(10) | 0.0310(27) | 1 |
| HV Vir | 0.057069(6) | 0.05821(5) | 0.0200(9) | 1, 14 |
| WZ Sge | 0.056687845 | 0.05721(4) | 0.0092(7) | 22 |
| DM Lyr | 0.06546(6) | 0.0673(2) | 0.0281(31) | 23, 24 |
| CU Vel | 0.0785(2) | 0.0808(2) | 0.0293(36) | 25, 26 |
| RX J2329+06 | 0.044567(4) | 0.04631(4) | 0.0391(9) | 27, 28, 29 |
| V359 Cen | 0.0779(3) | 0.08092(8) | 0.0388(40) | 30, 31 |
| XZ Eri | 0.06116(1) | 0.06281(10) | 0.0270(16) | 30, 32 |

[a]References: 1 = this work; 2 = Nogami, Kato, & Masuda 1998; 3 = Mennickent et al. 1999a; 5 = Nogami et al. 2000; 6 = Ishioka et al. 2001; 7 = Mennickent & Tappert 2001; 8 = Bond, Kemper, & Mattei 1982; 9 = Semeniuk et al. 1997; 10 = Kato 1996; 12 = Mennickent, Matsumoto, & Arenas 1999; 13 = Kato et al. 1998; 14 = Kato, Sekine, & Hirata 2001; 15 = Kato 1995; 16 = Thorstensen et al. 1996; 17 = Kato et al. 2001; 18 = Vrielmann & Offutt 2003; 19 = Kiyota & Kato 1998; 20 = Nogami et al. 1998; 21 = Kato et al. 2003; 22 = Patterson et al. 2002; 23 = Nogami et al. 2003; 24 = Thorstensen & Fenton 2002; 25 = Mennickent & Diaz 1996; 26 = Kato 2003; 27 = Thorstensen et al. 2002; 28 = Skillman et al. 2002; 29 = Uemura et al. 2002; 30 = Warner & Woudt 2001; 31=Kato et al. 2002; 32 = this work (in preparation).

# FIGURE CAPTIONS

FIGURE 1. — Averaged spectra of the ten stars. The vertical axes are in the units of $10^{-16}$ erg cm$^{-2}$ s$^{-1}$ Å$^{-1}$, but the flux scales are uncertain by at least 20 per cent.

FIGURE 2. — Period searches of the radial velocities (left panels), and radial velocities folded on the adopted periods (right panels). When data from several observing runs are combined, the many possible choices of cycle count between runs lead to fine-scale ringing in the periodogram. In these cases, the function plotted is formed by joining local maxima of the periodogram with straight lines, the word "peaks" appears in the title, and the period used in folding the velocities for the right-hand panel reflects an arbitrary choice of cycle count between observing runs. Two cycles are shown in the folds for continuity.

FIGURE 3. — See Figure 2 caption.

FIGURE 4. — Phase-averaged spectra of four stars, showing faint *S*-waves in the He I lines.

FIGURE 5. — Power spectra of light curves of nine dwarf novae during supermaximum, with one star (KS UMa) repeated since it showed a significant, and unusual, difference between eruptions. These (plus TU Crt) are the stars for which we have spectroscopy at quiescence — the stars of Figures 1 – 3. Flagged frequencies are typically accurate to ±0.02 c/d, but see text for more detailed error discussion. In a few cases the frequencies differ slightly from those adopted in the text or table; this is because the harmonic was also used, or because of slight variability.

FIGURE 6. — *Upper frame*, power spectrum of RZ Leo in quiescence. *Lower left frame*, O–C diagram of the nightly timings of maximum light in quiescence, demonstrating the stability of the period. *Lower right frame*, mean quiescent light curve.

FIGURE 7. — BC UMa in superoutburst, February/March 2000. *Upper frame*, power spectrum of the first four days, with a possible signal at 31.97(4) c/d. Inset is the mean waveform (summed at $P_o$). *Lower frame*, light curve during the next four days, showing the rapid growth of common superhumps.

FIGURE 8. — Power spectrum of BC UMa in quiescence, with a signal at 31.92(5) c/d likely to signify $\omega_o$. Inset is the mean orbital light curve.

FIGURE 9. — Power spectrum of LL And in quiescence (*V*=20), showing a strong signal at 18.163(3) c/d which we interpret as $\omega_o$. Aliases at 18.124 and 18.203 c/d are possible, but unlikely. Inset is the mean light curve at quiescence.

FIGURE 10. — Power spectra of KV And in superoutburst and quiescence (*V*~19.5); inset is the mean quiescent light curve.

FIGURE 11. — Power spectrum of WX Cet in the November 1998 superoutburst.

FIGURE 12. — Power spectrum of WX Cet on 1996 July 23, showing a dwarf-nova oscillation at 17.385(6) s.

FIGURE 13. — *Upper frame*, power spectrum of MM Hya in superoutburst. *Second frame*, power spectrum near quiescence. *Third frame*, cleaned quiescent power spectrum, showing the closely spaced frequencies ($\omega_{sh}$ and $\omega_o$). *Bottom frame*, the mean light curve and O–C diagram at quiescence.

FIGURE 14. — Power spectra of AO Oct in superoutburst and quiescence. Inset is the quiescent light curve.

FIGURE 15. — Power spectra of V2051 Oph in the 1998 and 1999 superoutbursts. See text for a tortuous discussion of frequencies, culminating in $\omega_{sh}$=15.58(4) c/d.

FIGURE 16. — Power spectra of NY Ser in superoutburst and quiescence. Inset is the mean quiescent light curve.

FIGURE 17. — Power spectra of KK Tel in superoutburst and quiescence.

FIGURE 18. — The upper two frames show the superoutburst and quiescent power spectra of HV Vir. The lowest frame shows a close-up of the region around $2\omega_o$ (based on the full 58-day time series), and the mean orbital light curve.

FIGURE 19. — *Upper frame*, power spectrum of RX J1155.4–5641 in superoutburst. *Lower left frame*, portion of the power spectrum of residuals after the superhump is removed from the time series. *Lower right frame*, portion of the power spectrum near quiescence. The signals in the lower frames suggest that $\omega_o$ may be 16.59(2) c/d.

FIGURE 20. — Distribution of $\epsilon$ with $P_o$ for H-rich apsidal superhumpers. The solid curve is the trend predicted if the secondaries are BCAH ZAMS stars ($\alpha$=1). The dashed extension at short $P_o$ shows the predicted effect of disequilibrium in theoretical BCAH stars, assuming it arises from angular momentum loss due to gravitational radiation (GR). The disagreement is obvious, but the points appear to define a curve *similar* to the theoretical curve.

FIGURE 21. — The shaded region shows the values of $\alpha$ and $<M_1>$ which satisfy the superhump data on 46 stars in the best-constrained $P_o$ range (0.060–0.085 d). The boxes show the measured values of $\alpha$ and $M_1$ obtained for the six stars (eclipsing binaries) with accurately measured mass and radius. An average $M_1$ of 0.75 $M_\odot$ appears to be appropriate for these stars, as well as for CVs generally (see text). This implies $\alpha$=1.18(6).

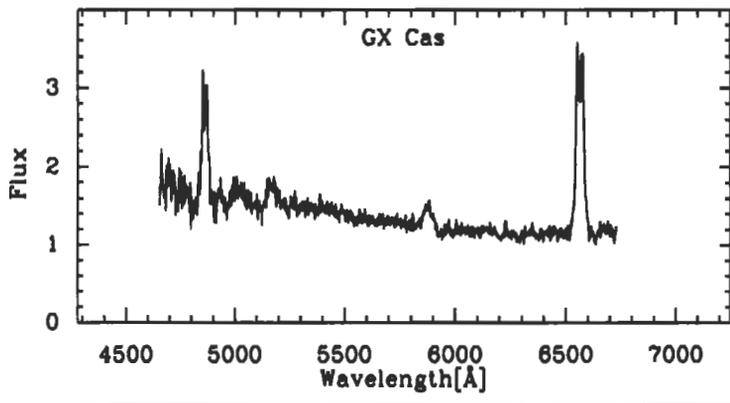
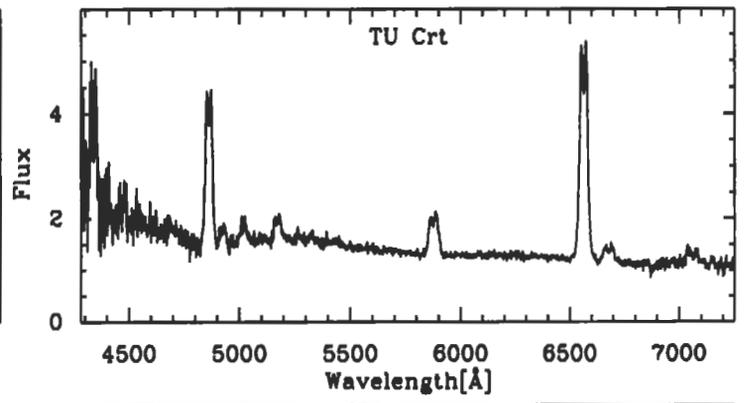
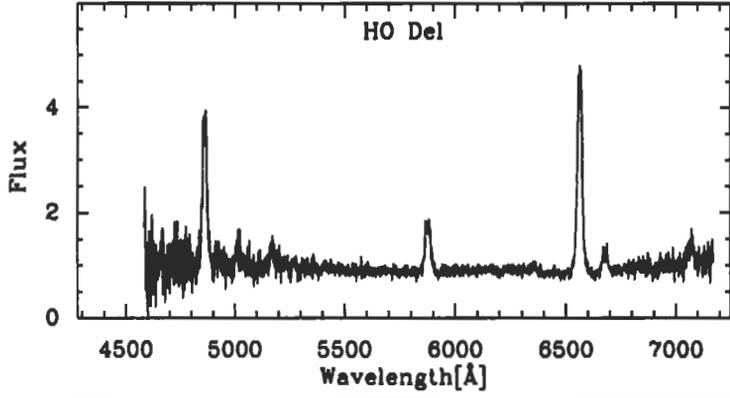
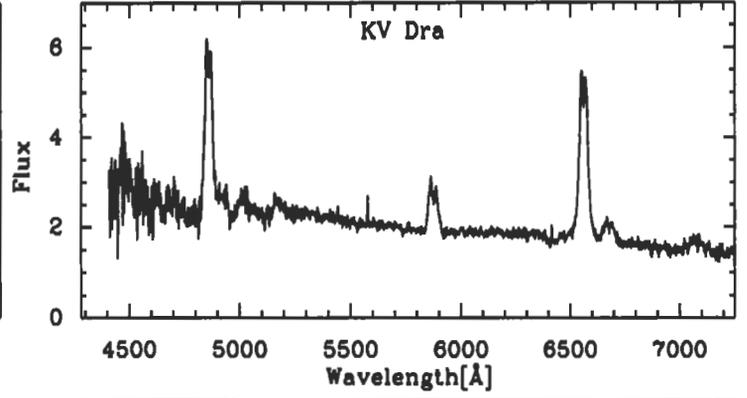
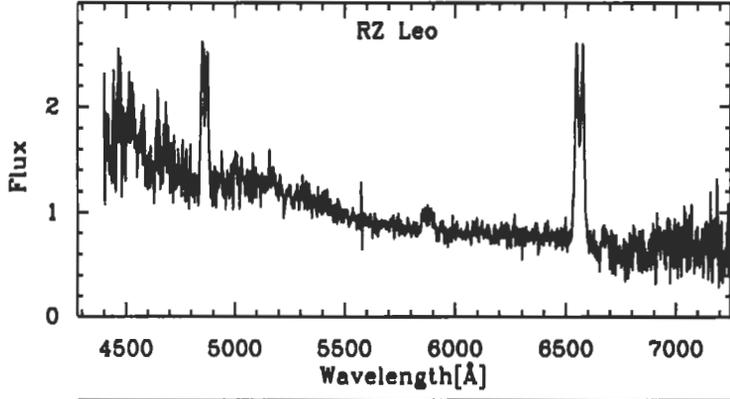
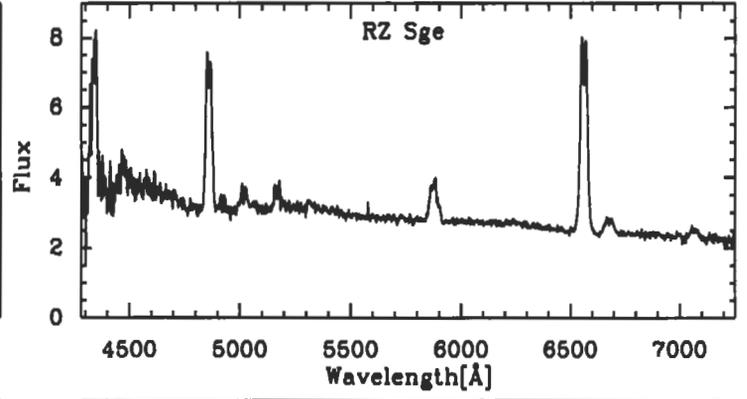
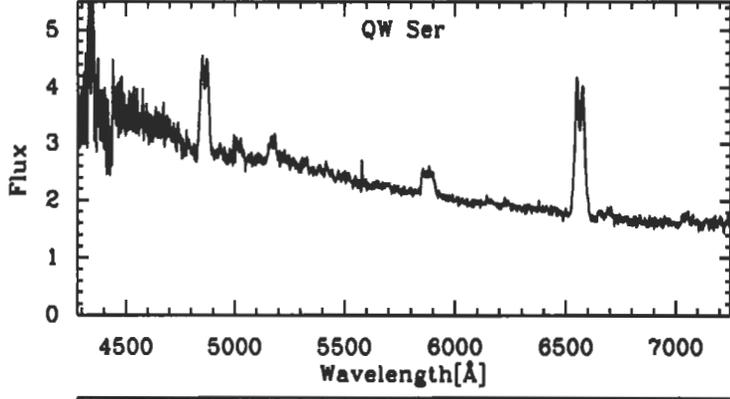
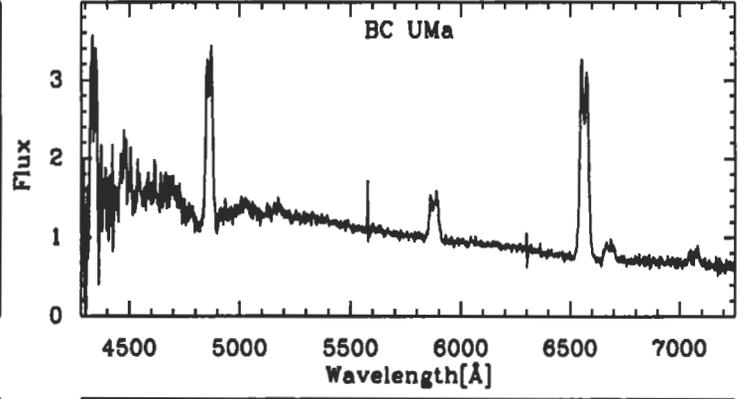
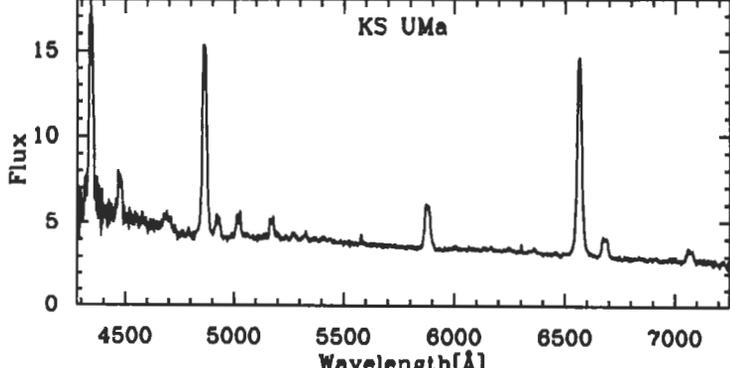
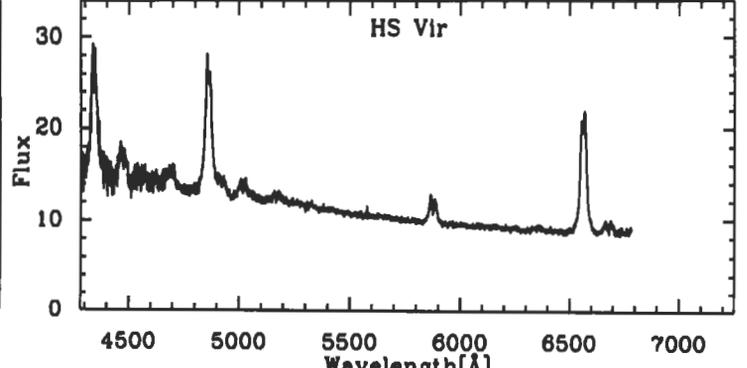

Fig 1

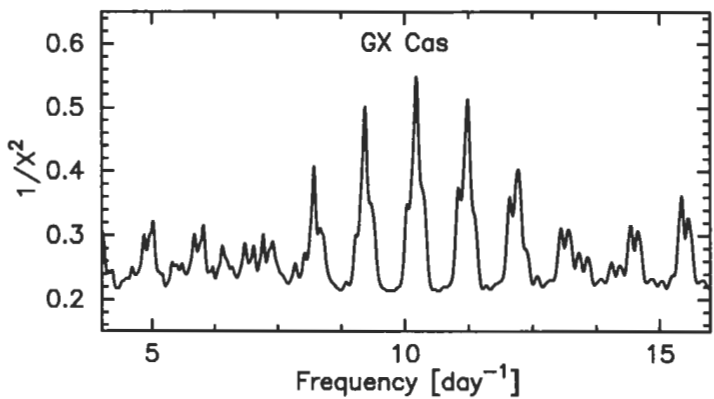
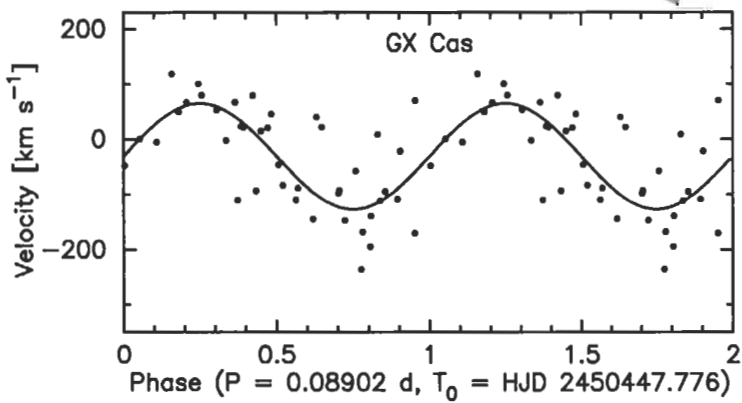
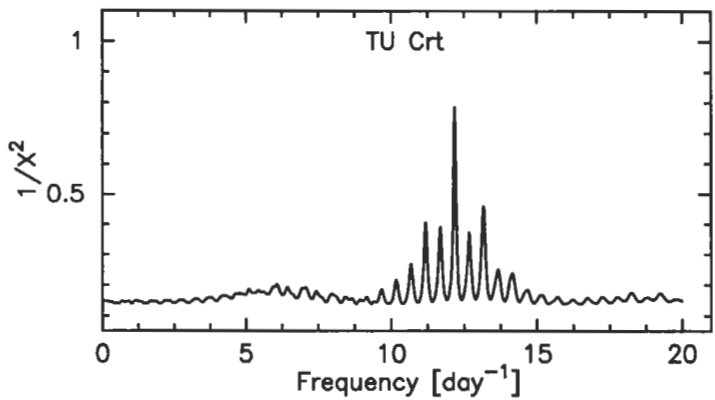
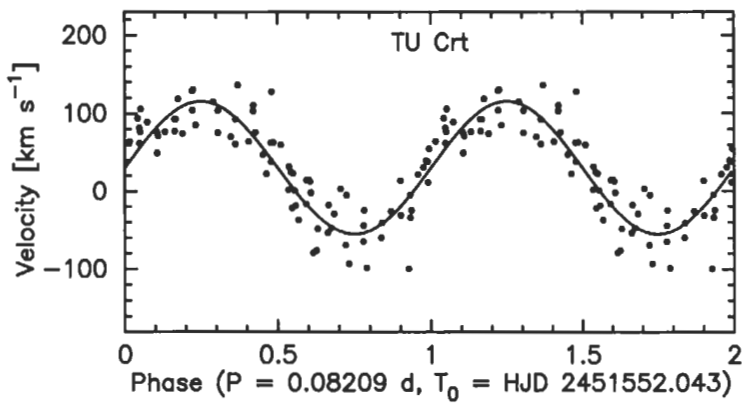
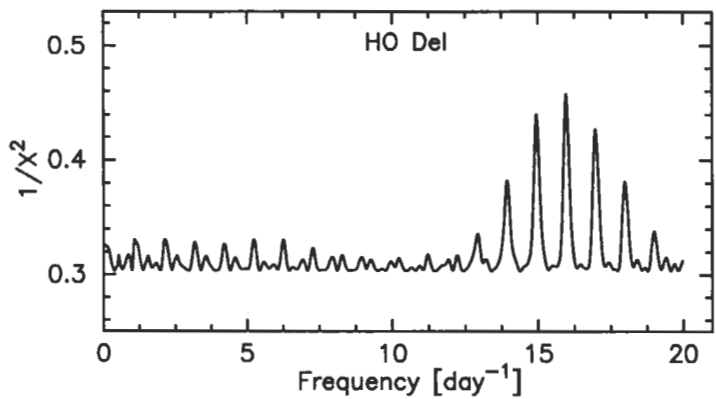
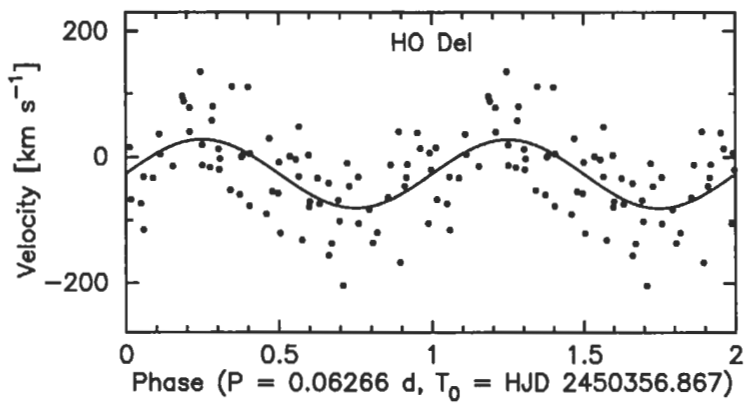
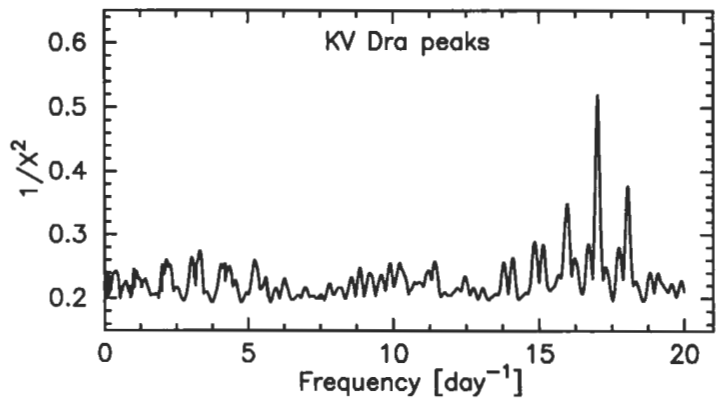
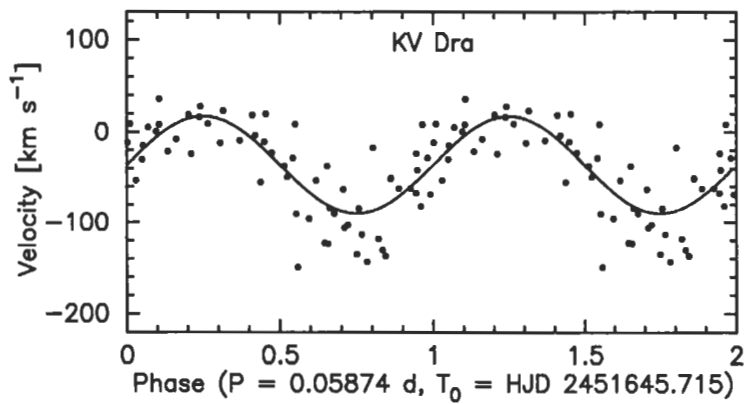
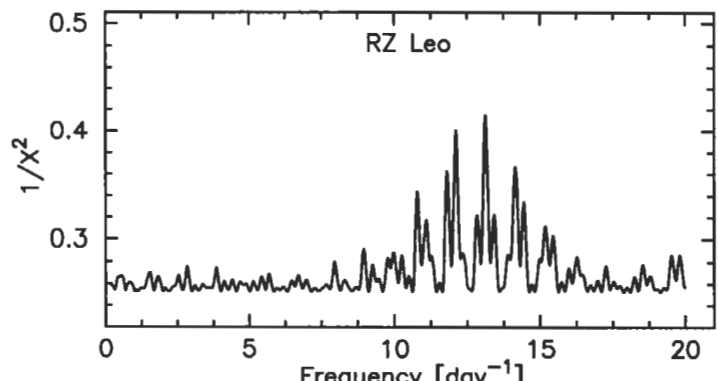
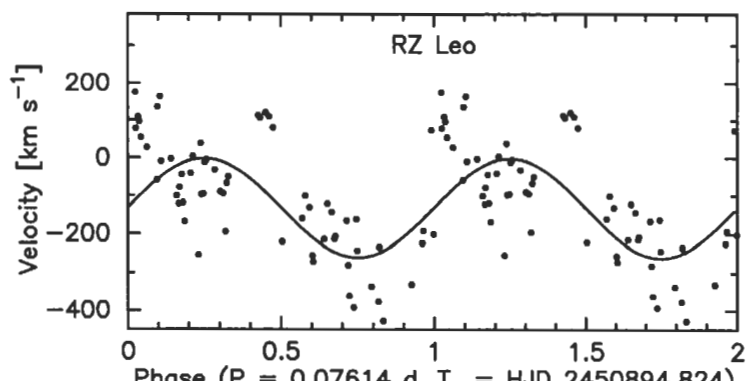

Fig 2

Fig 5

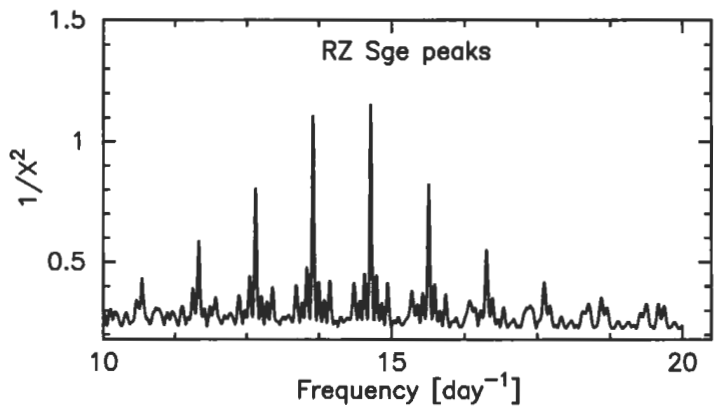
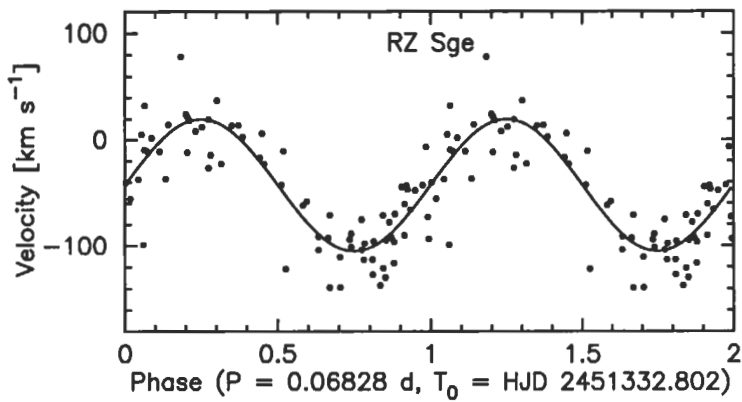
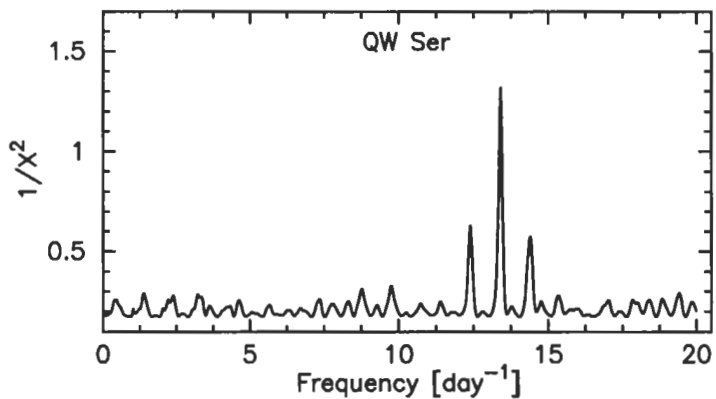
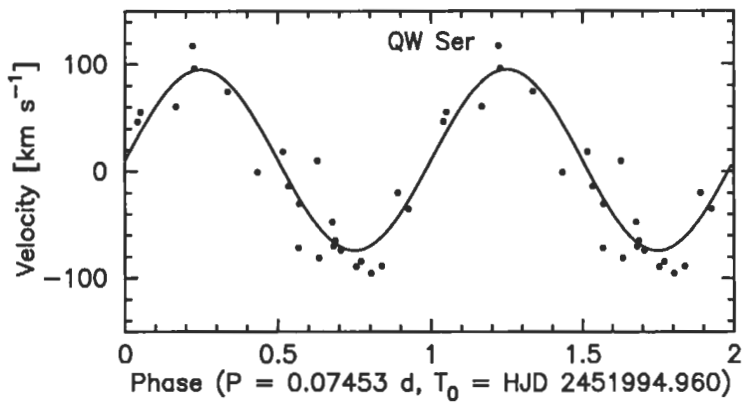
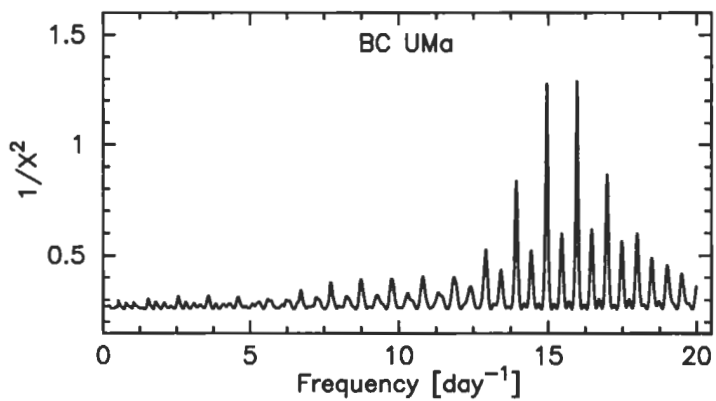
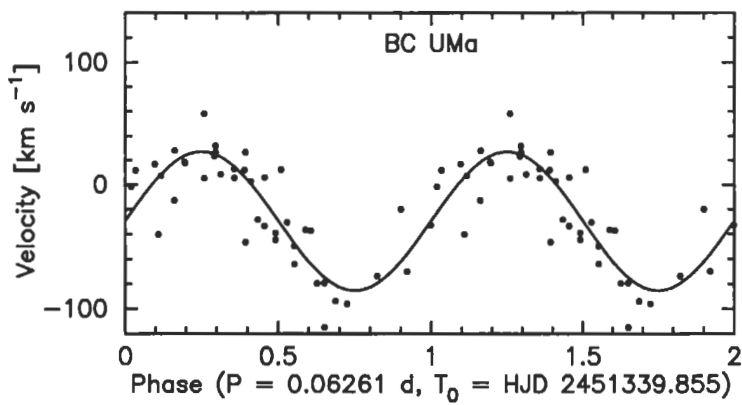
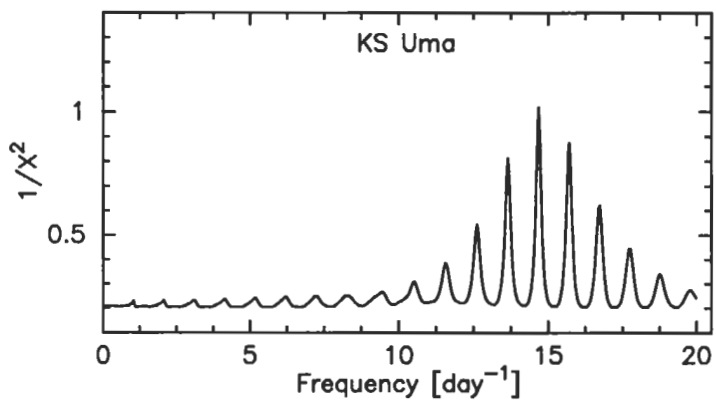
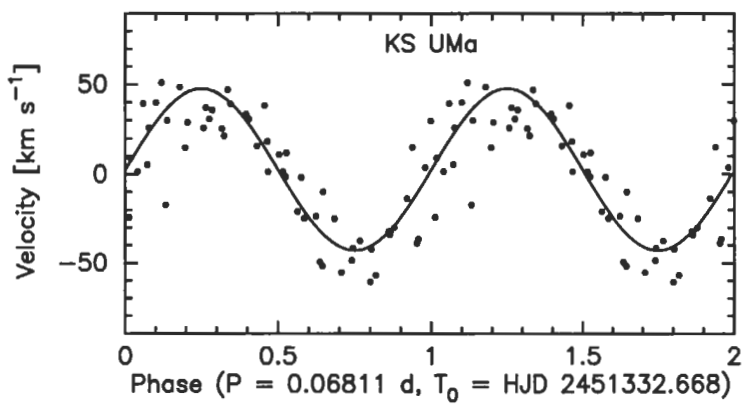
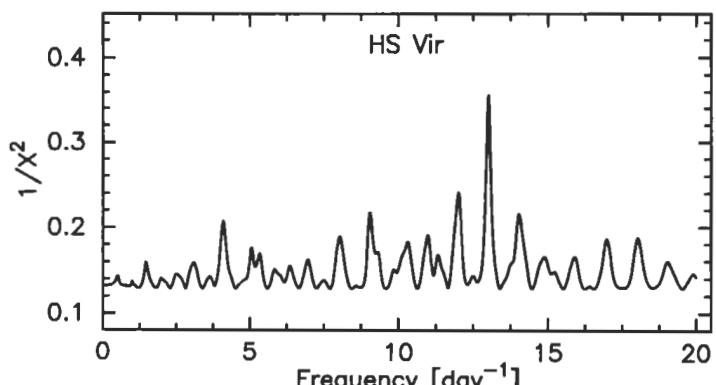
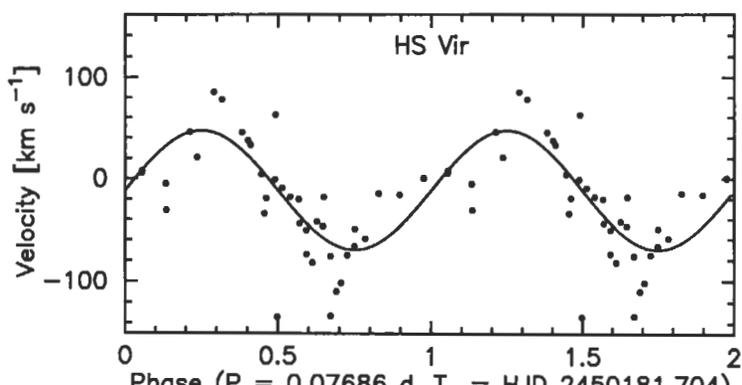

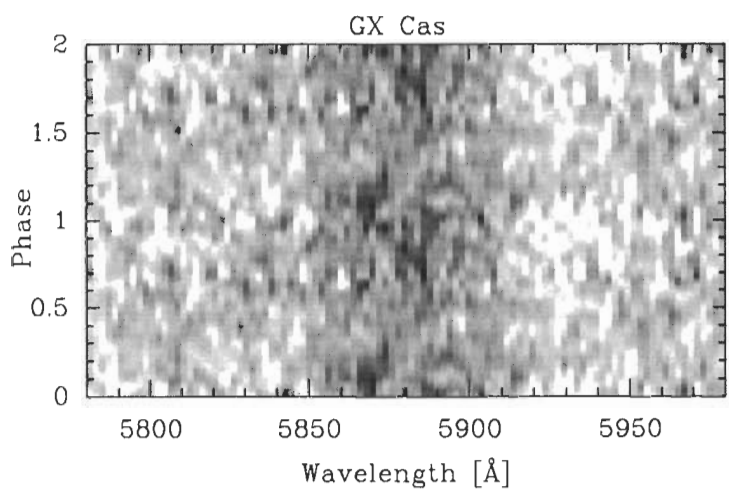
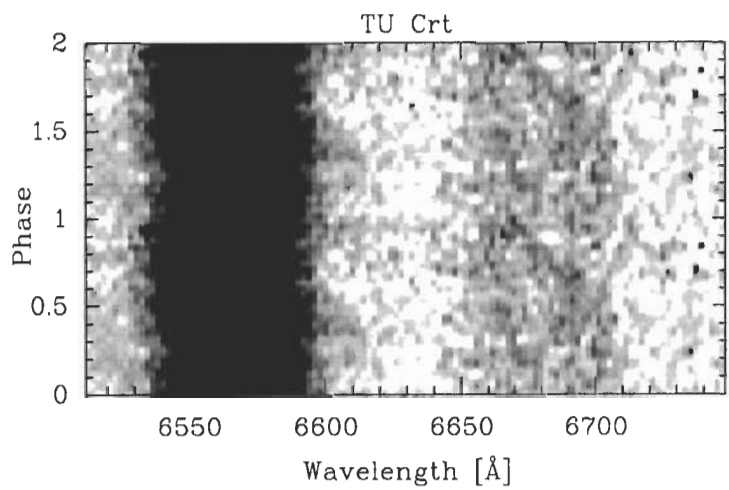
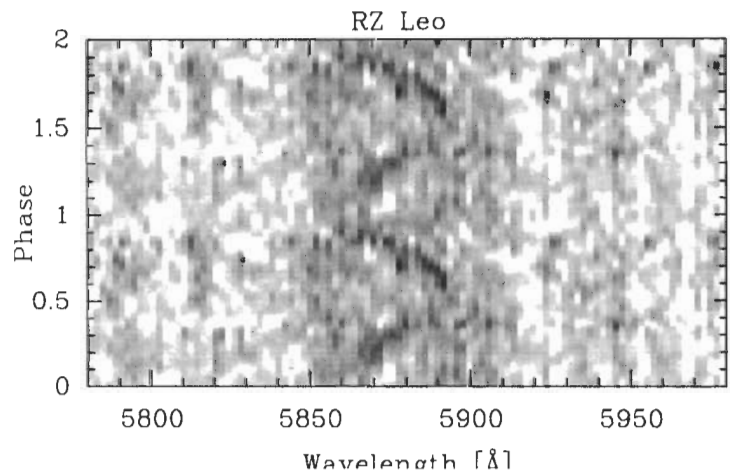
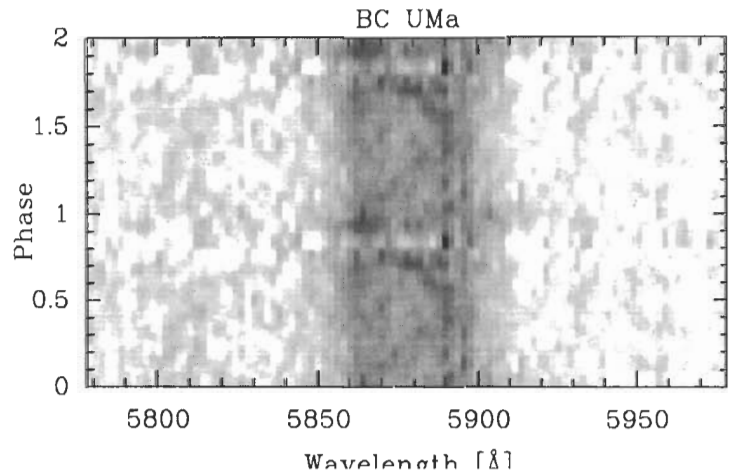

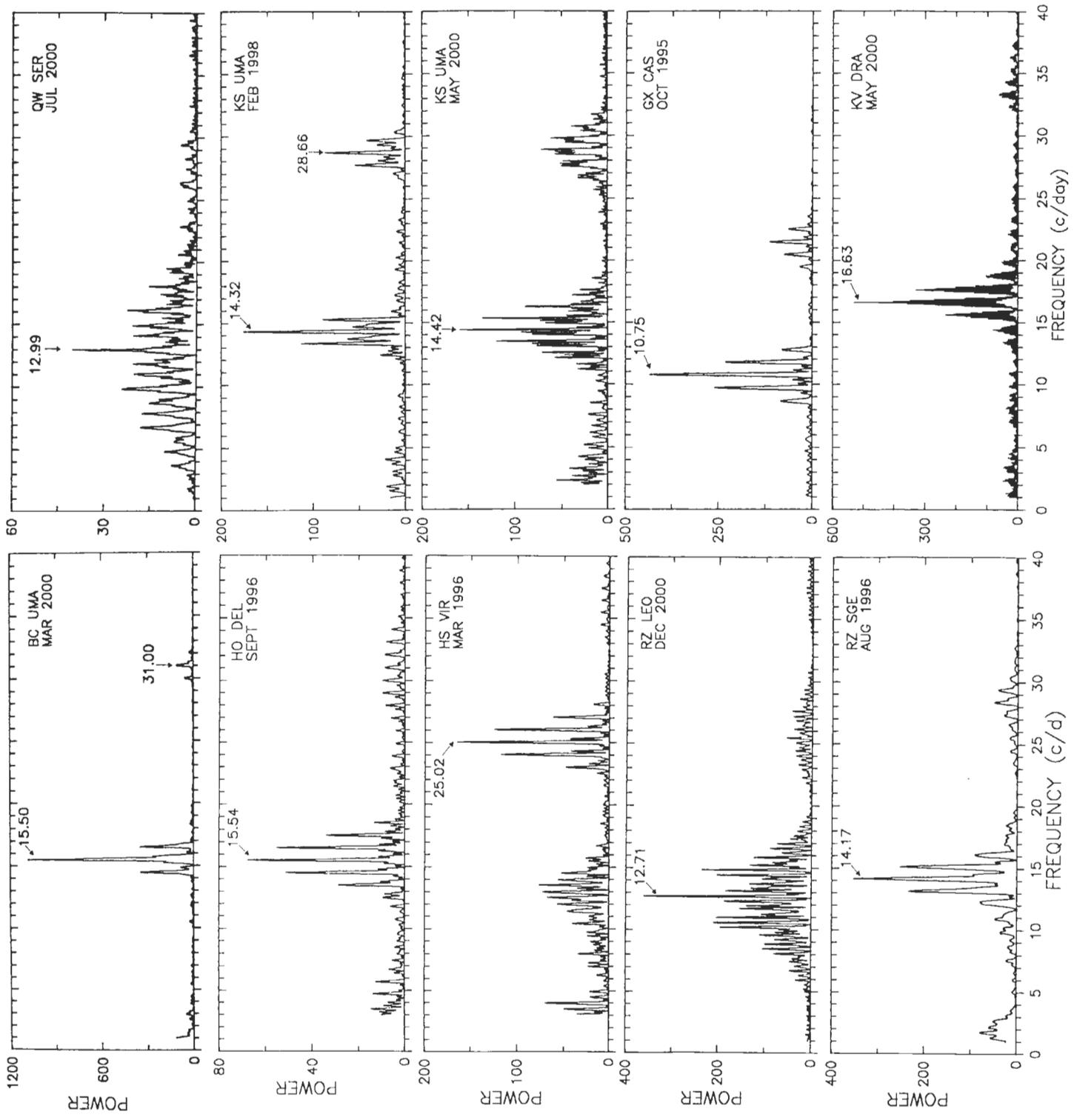

Fig 5

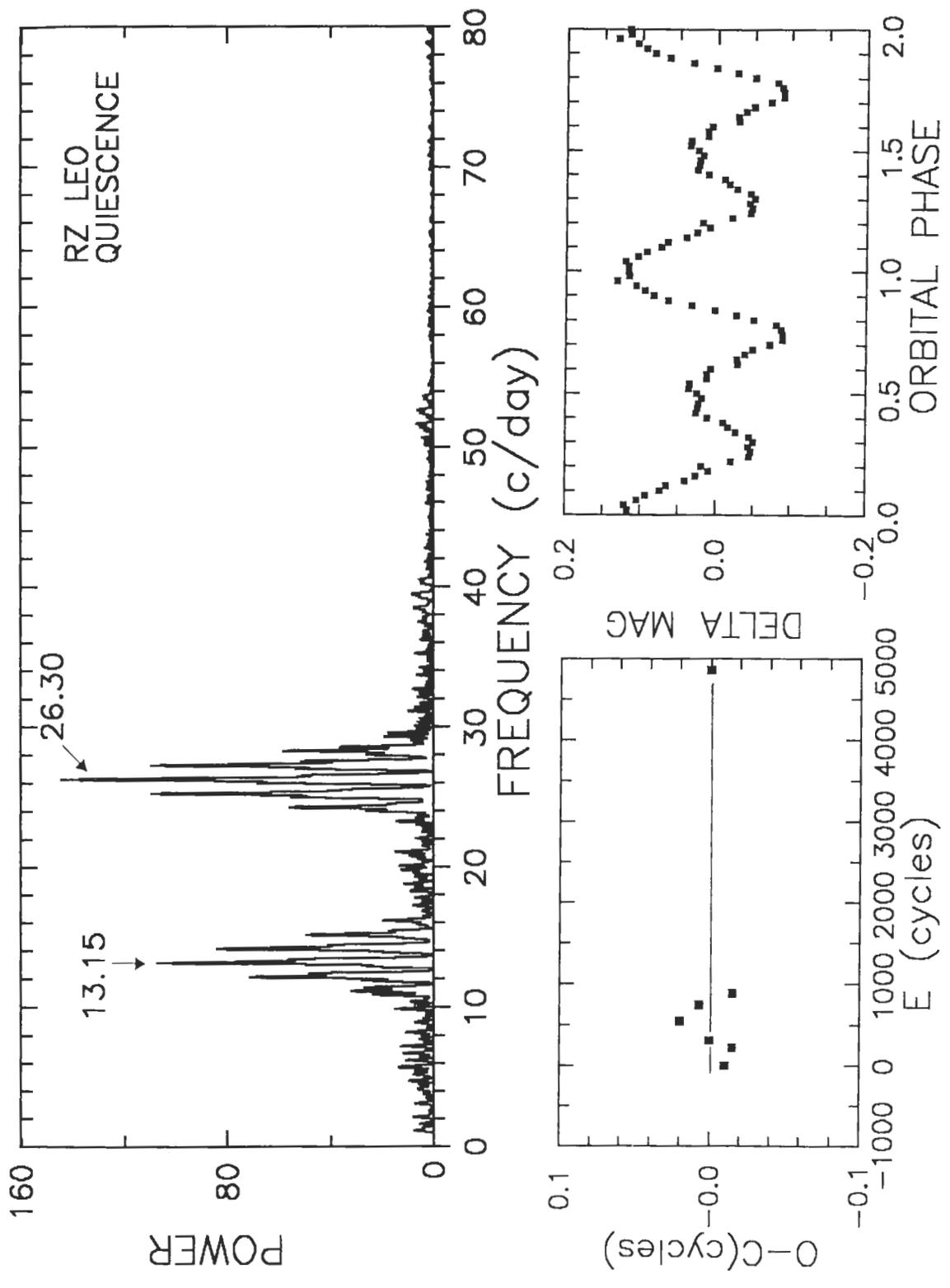

Fig 6

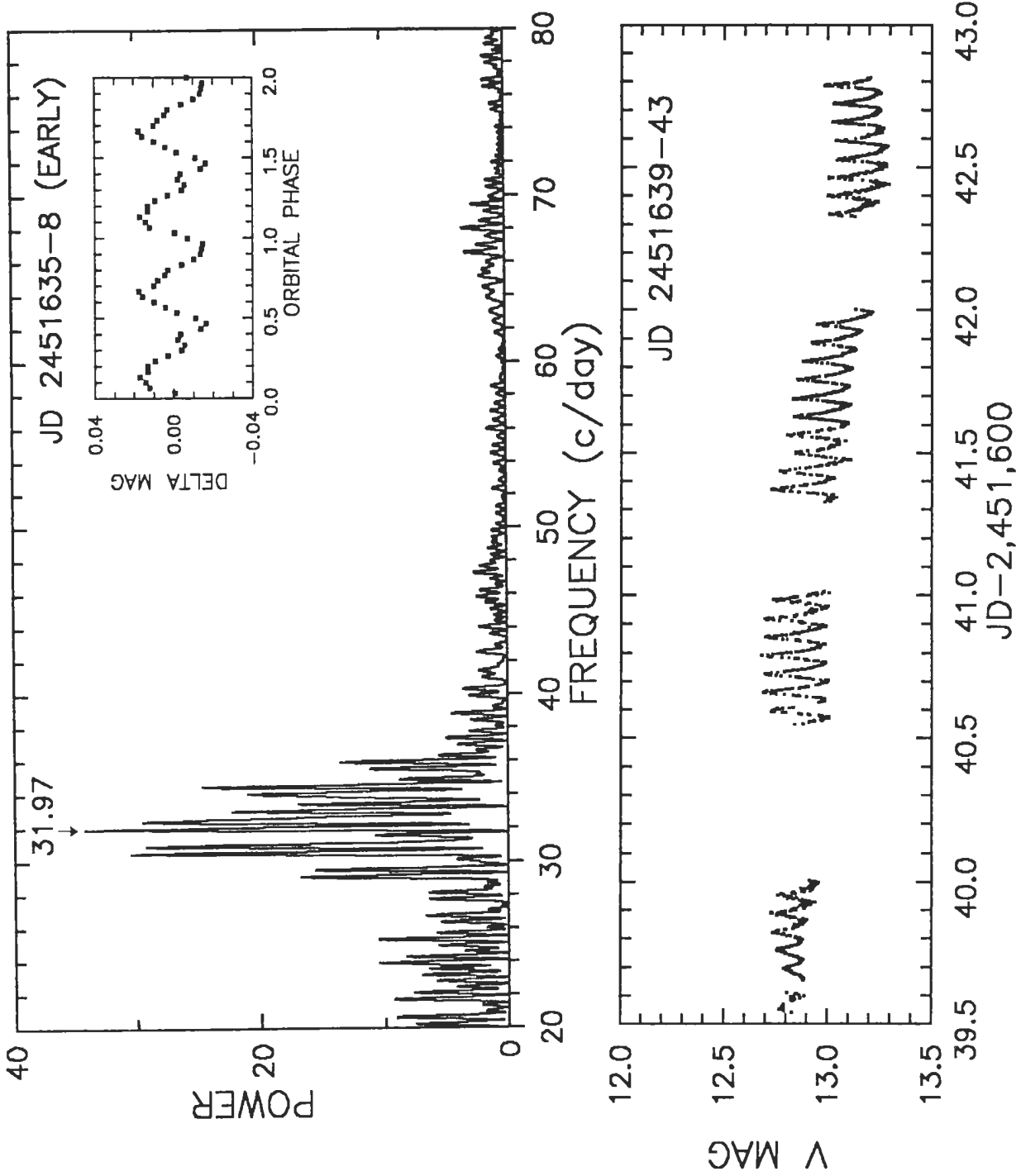

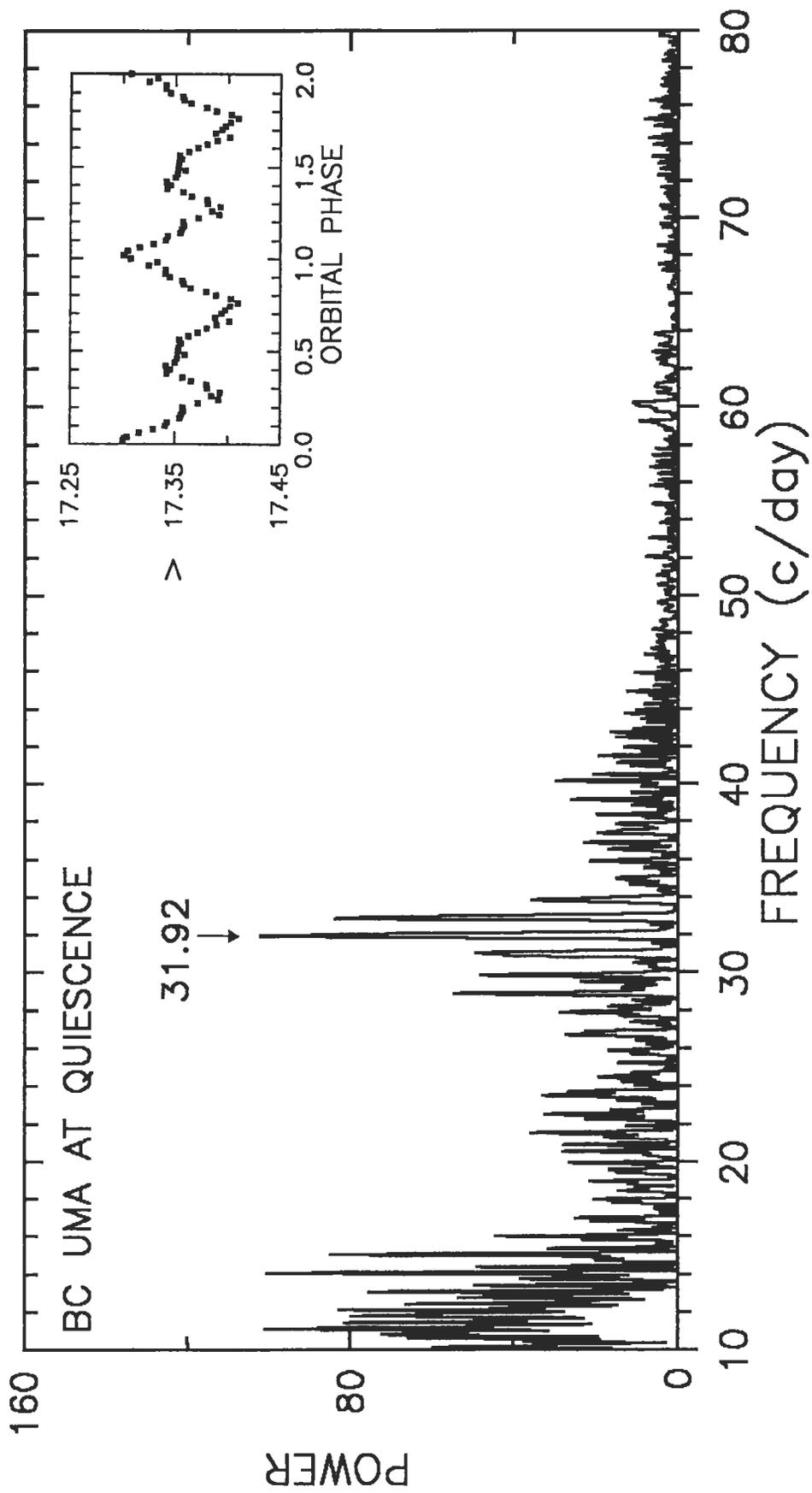

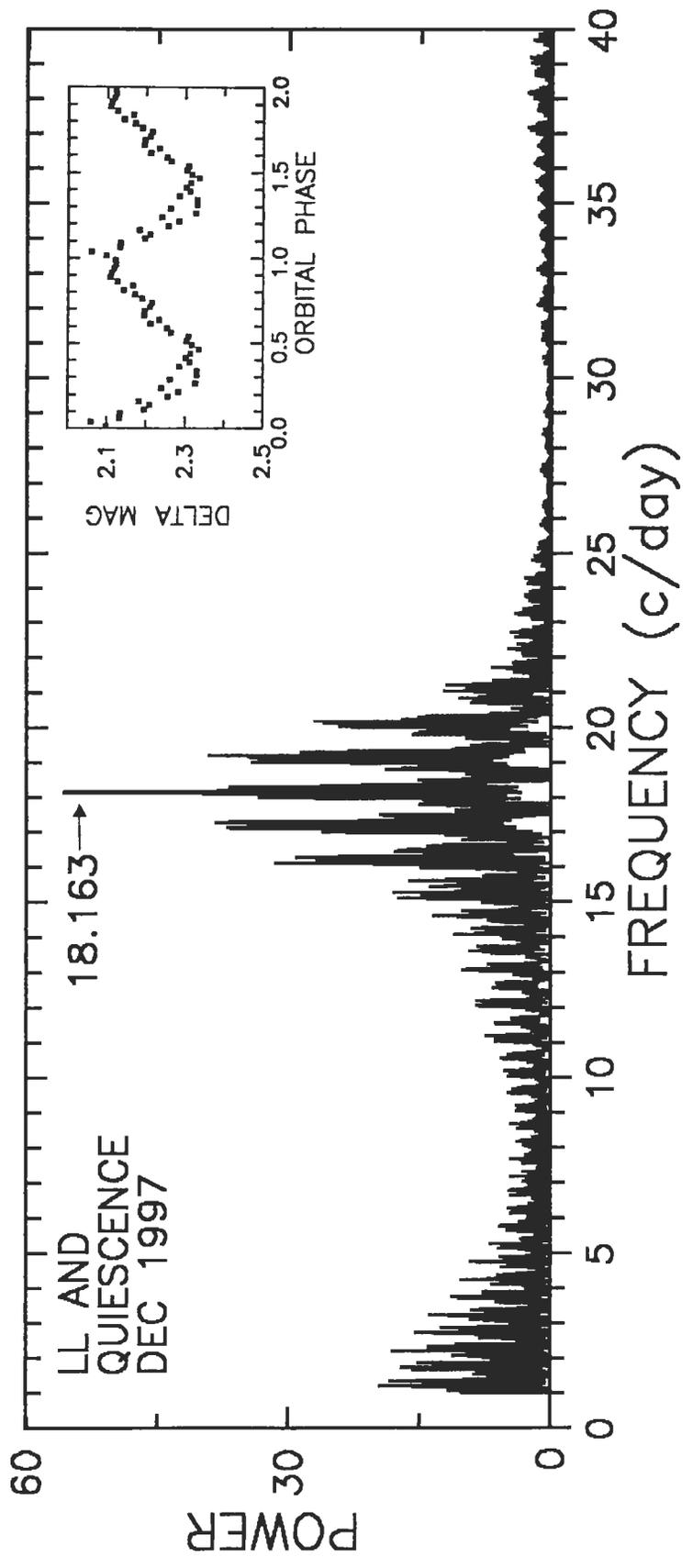

Fig 9

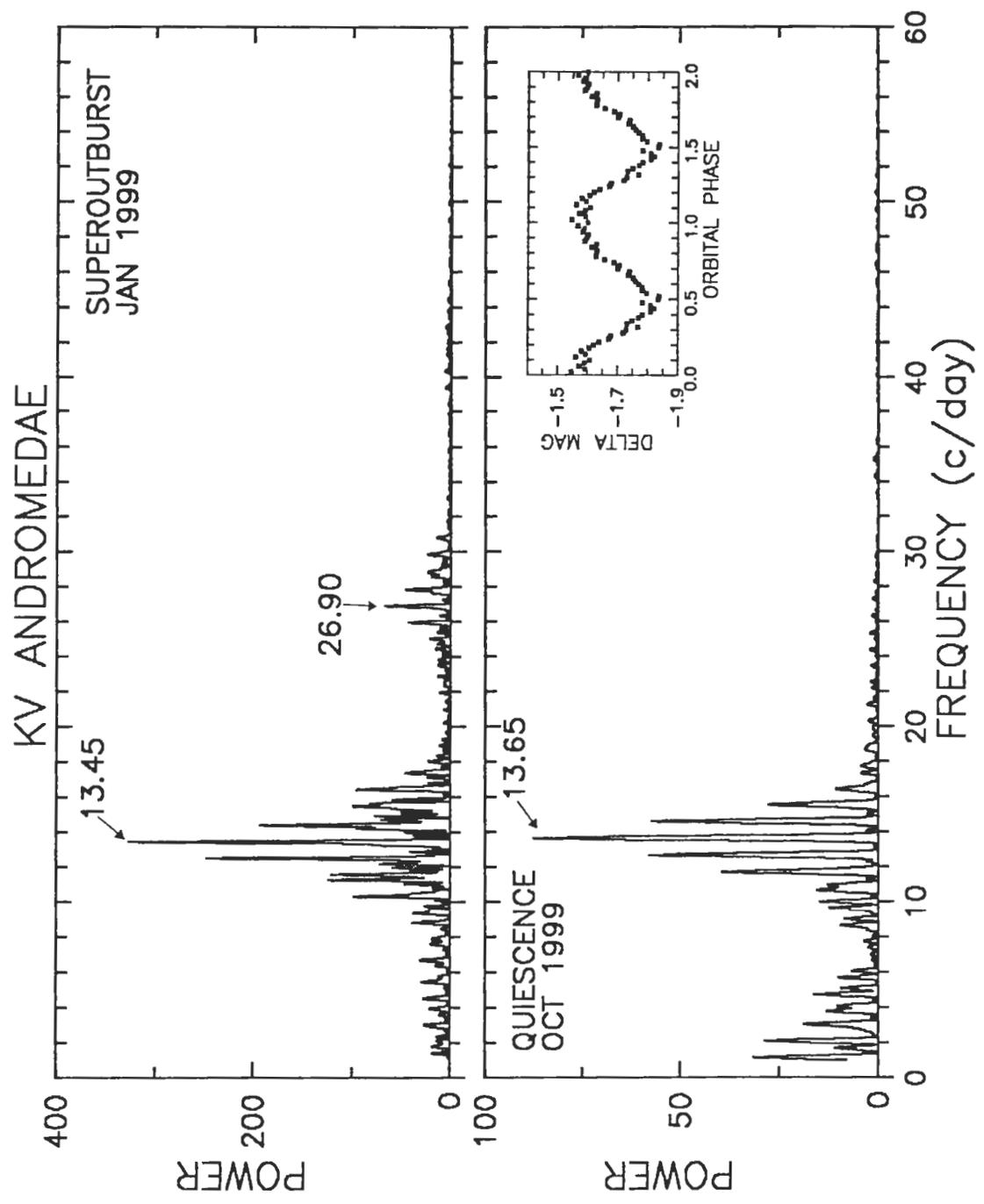

Fig 10

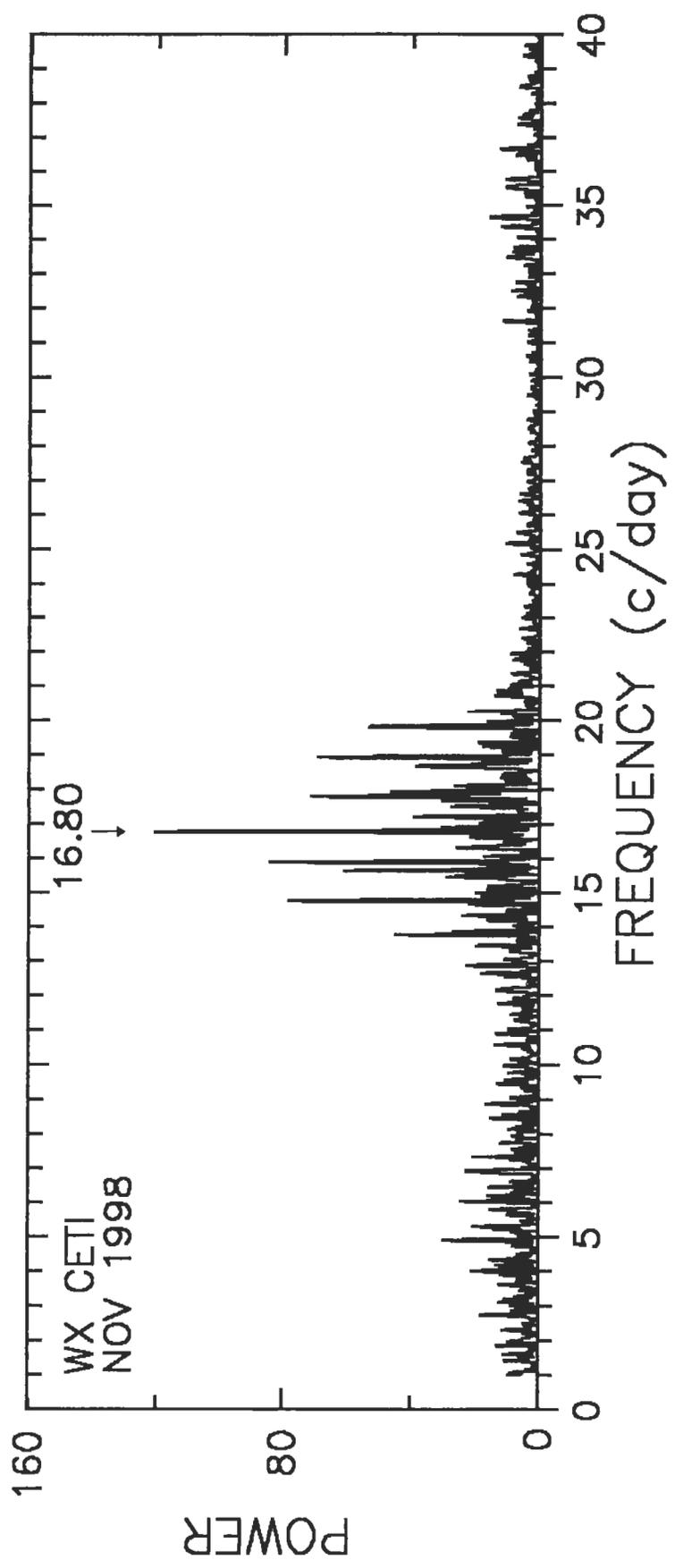

Fig 11

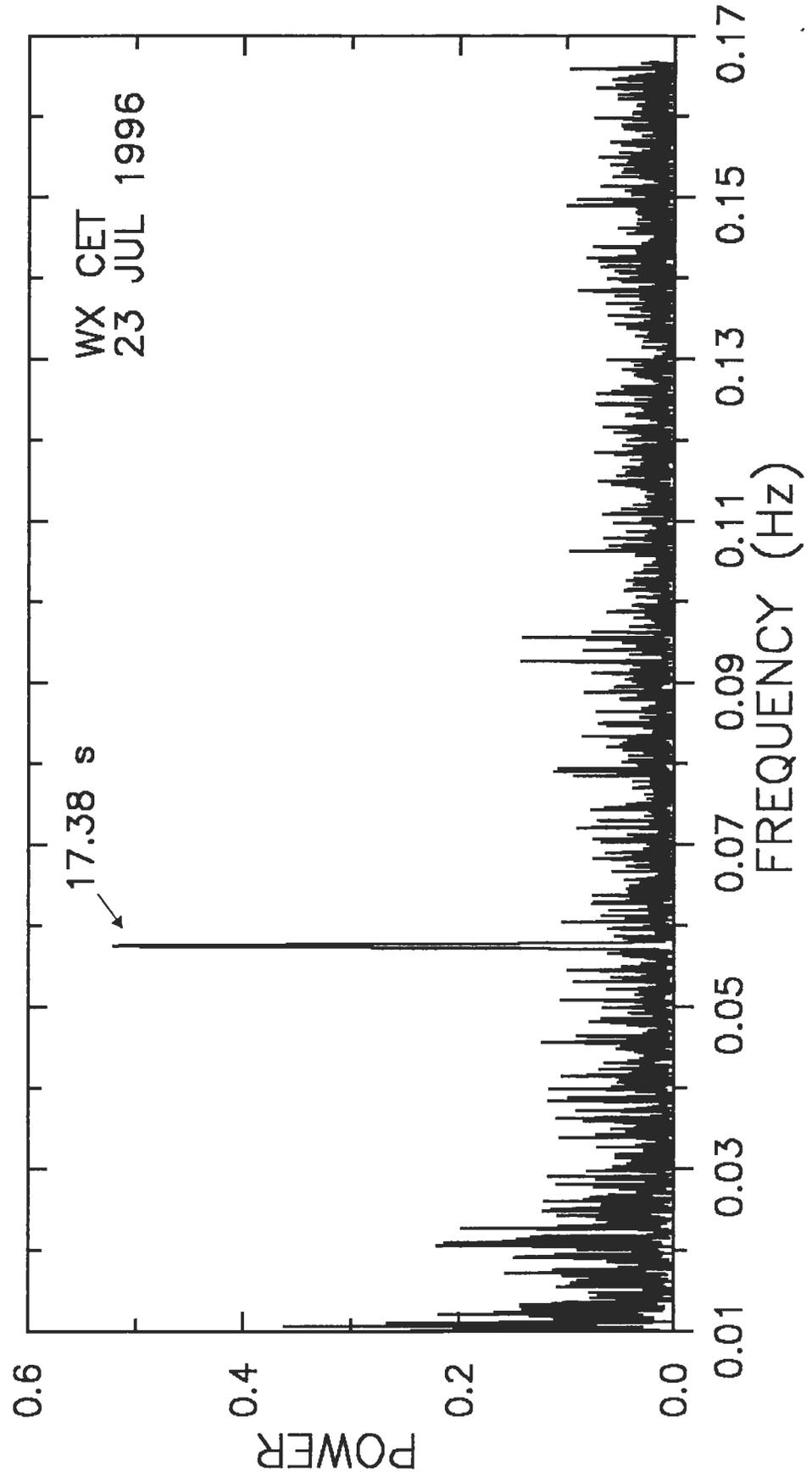

Fig 12

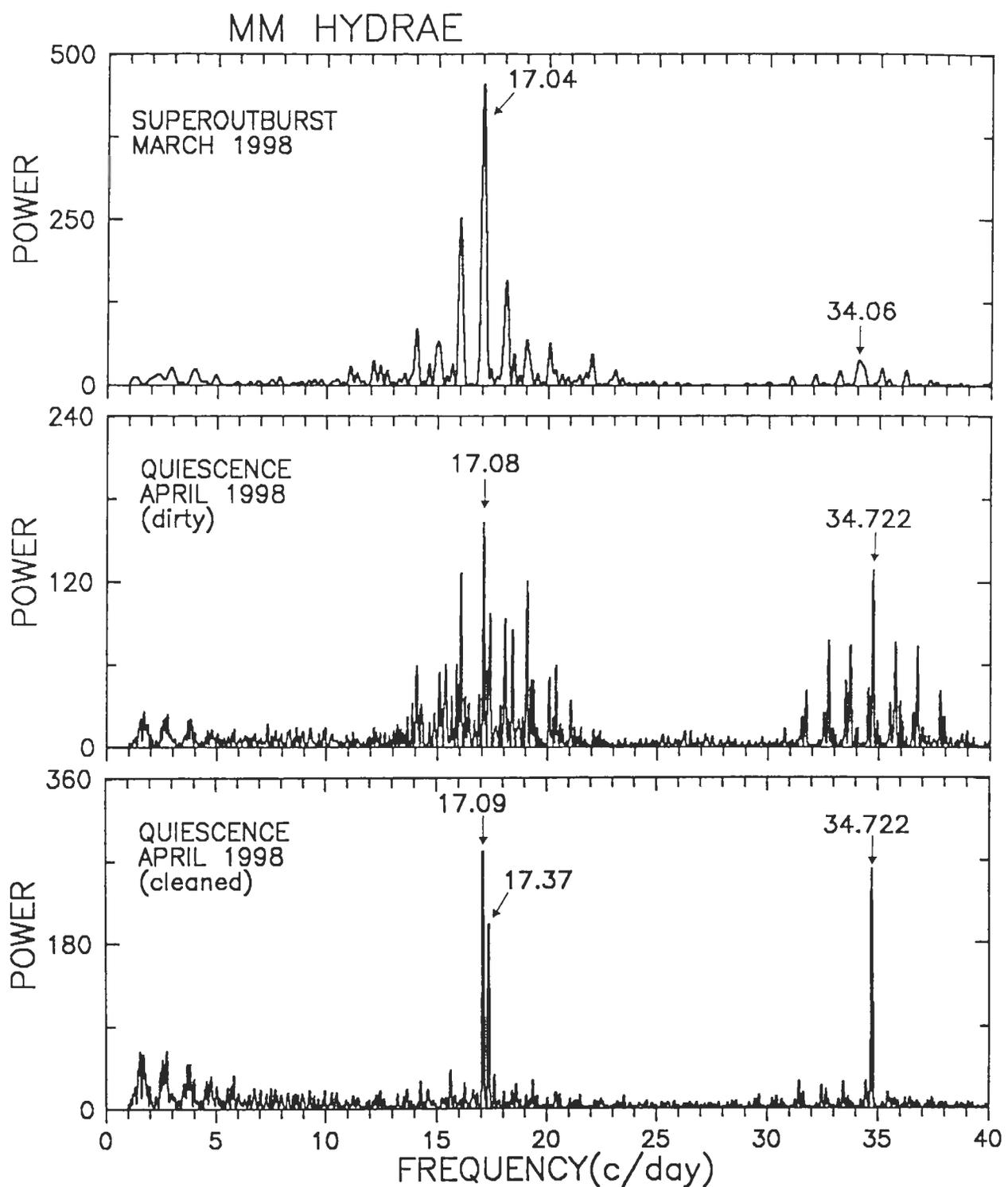
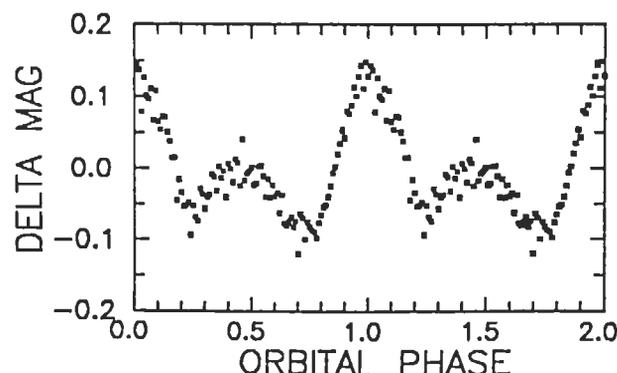
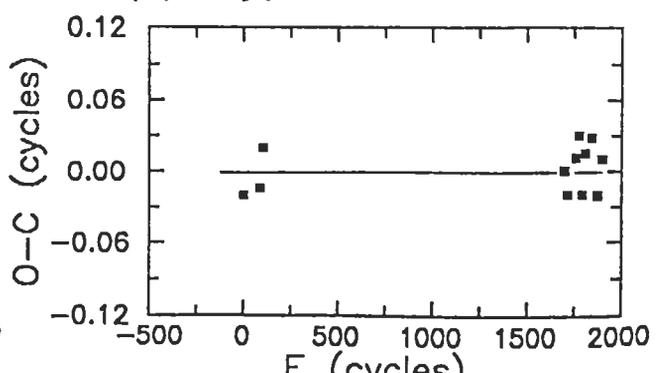
Fig 13

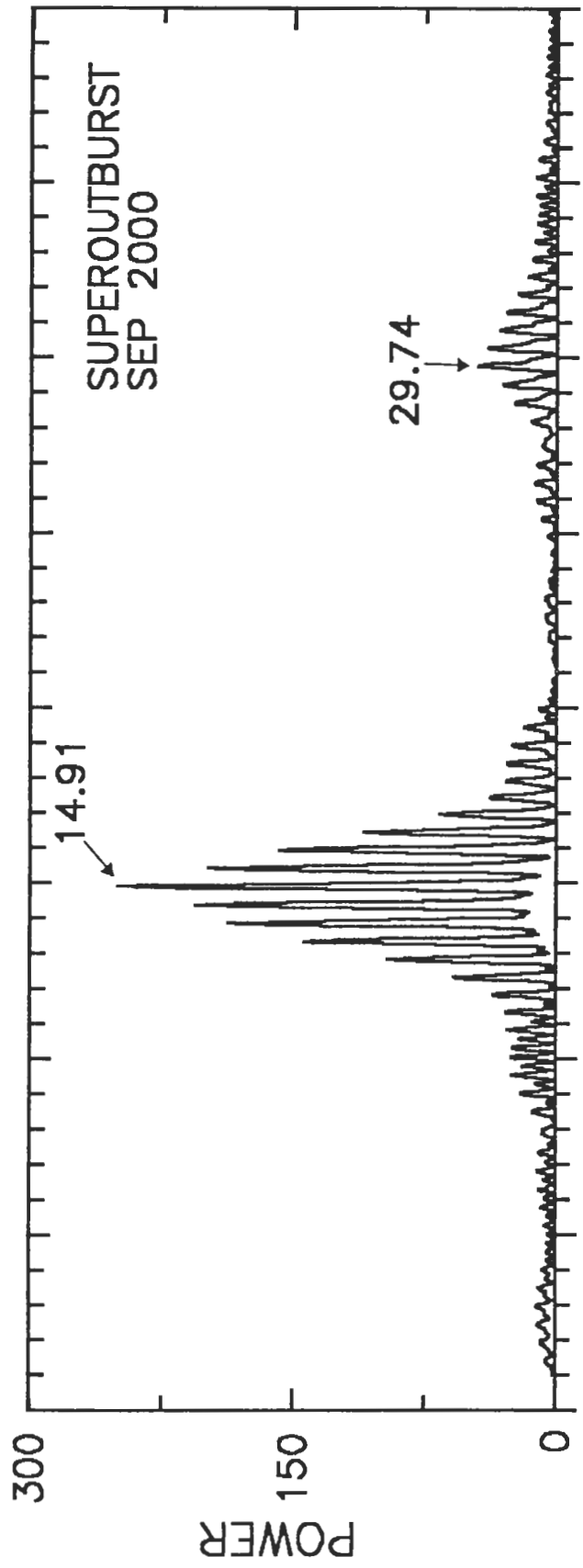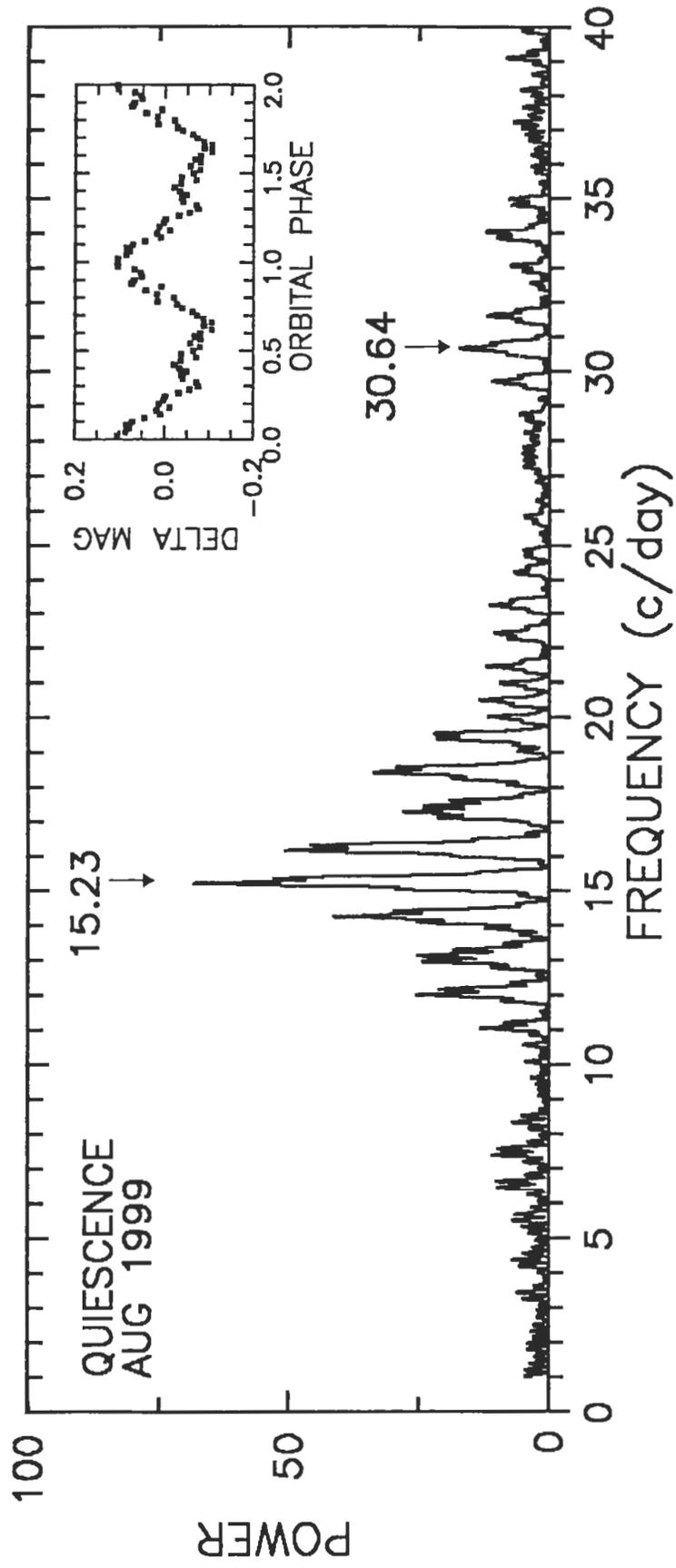

Fig 14

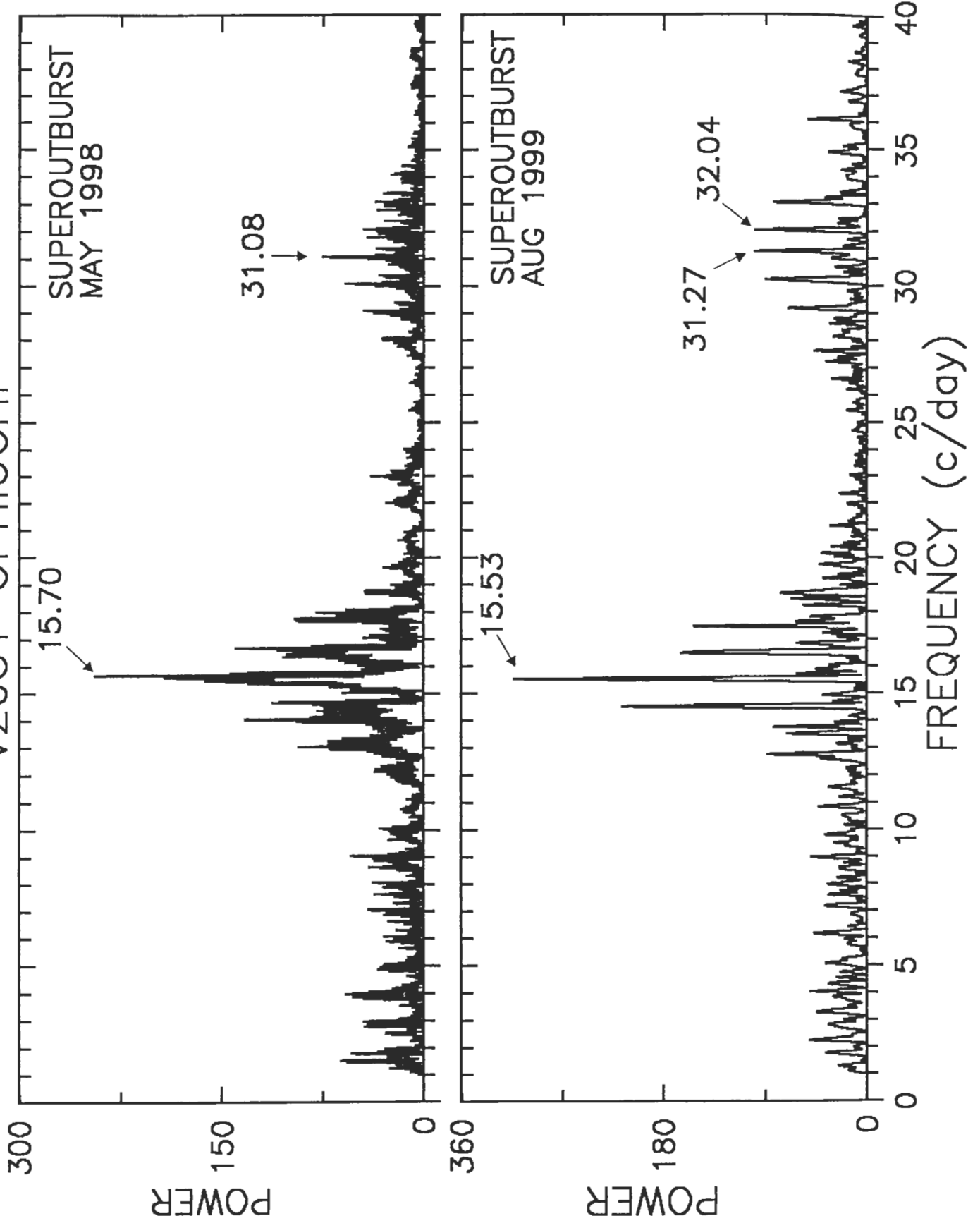

Fig 15

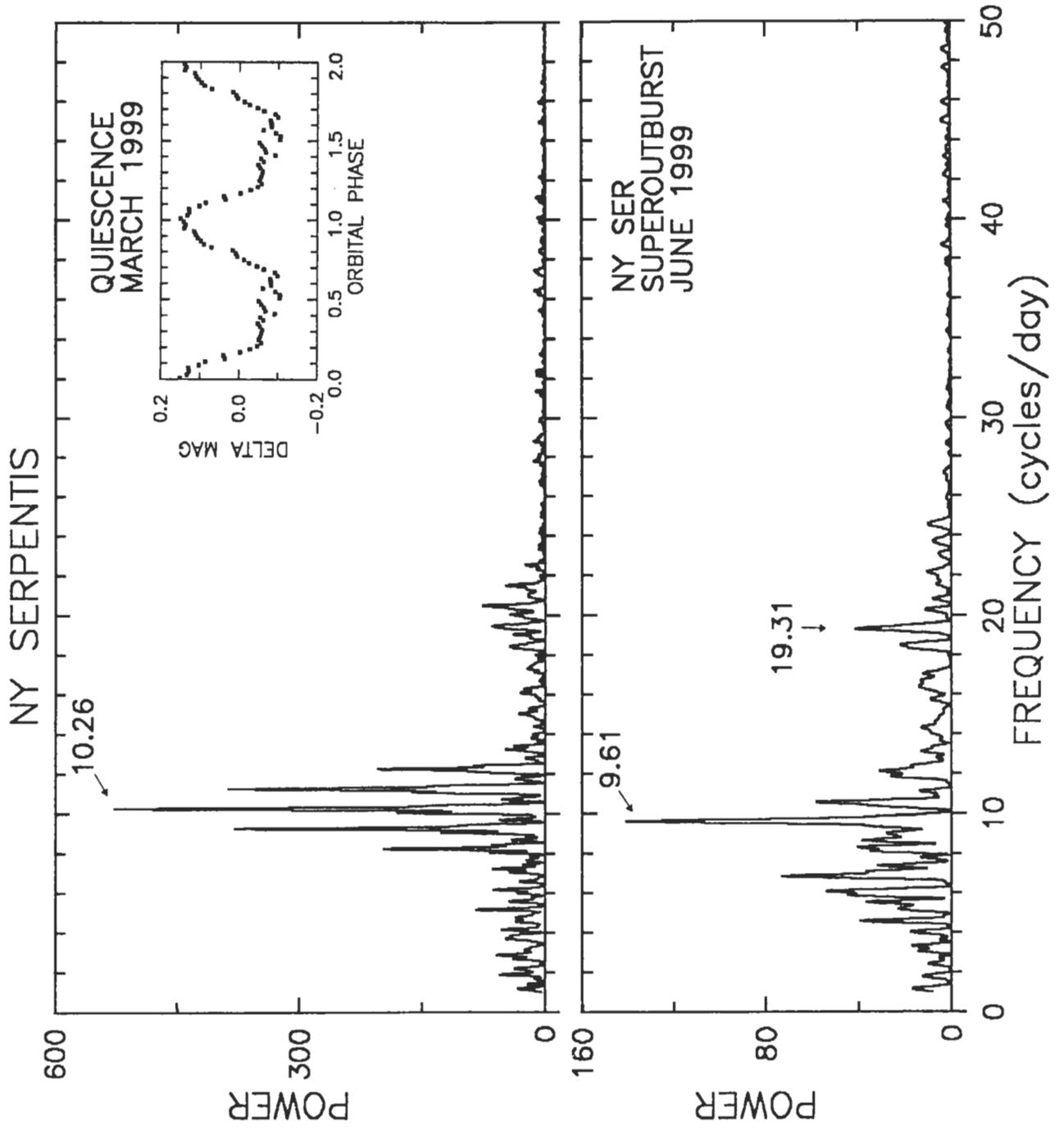

Fig 16

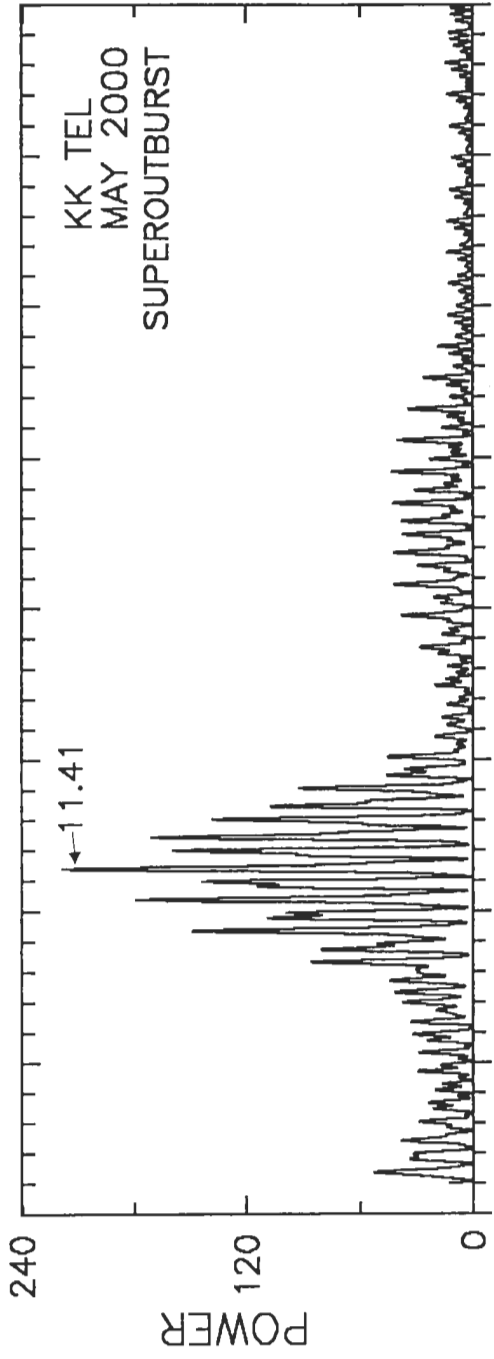
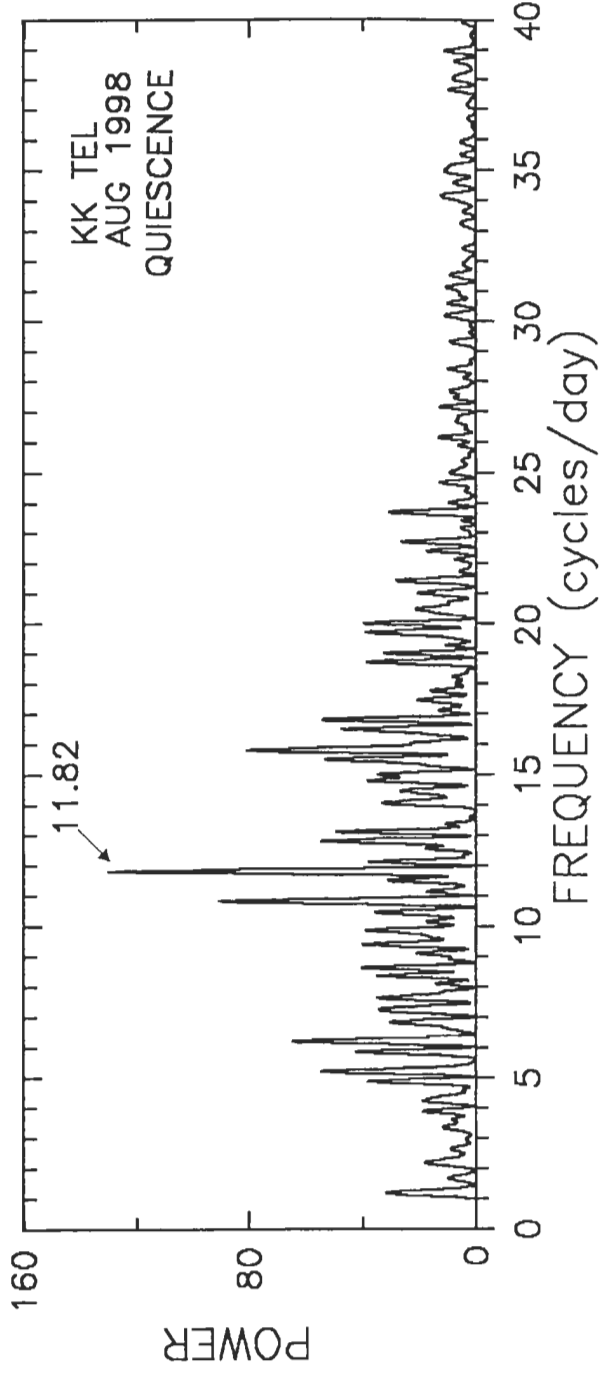

Fig 17

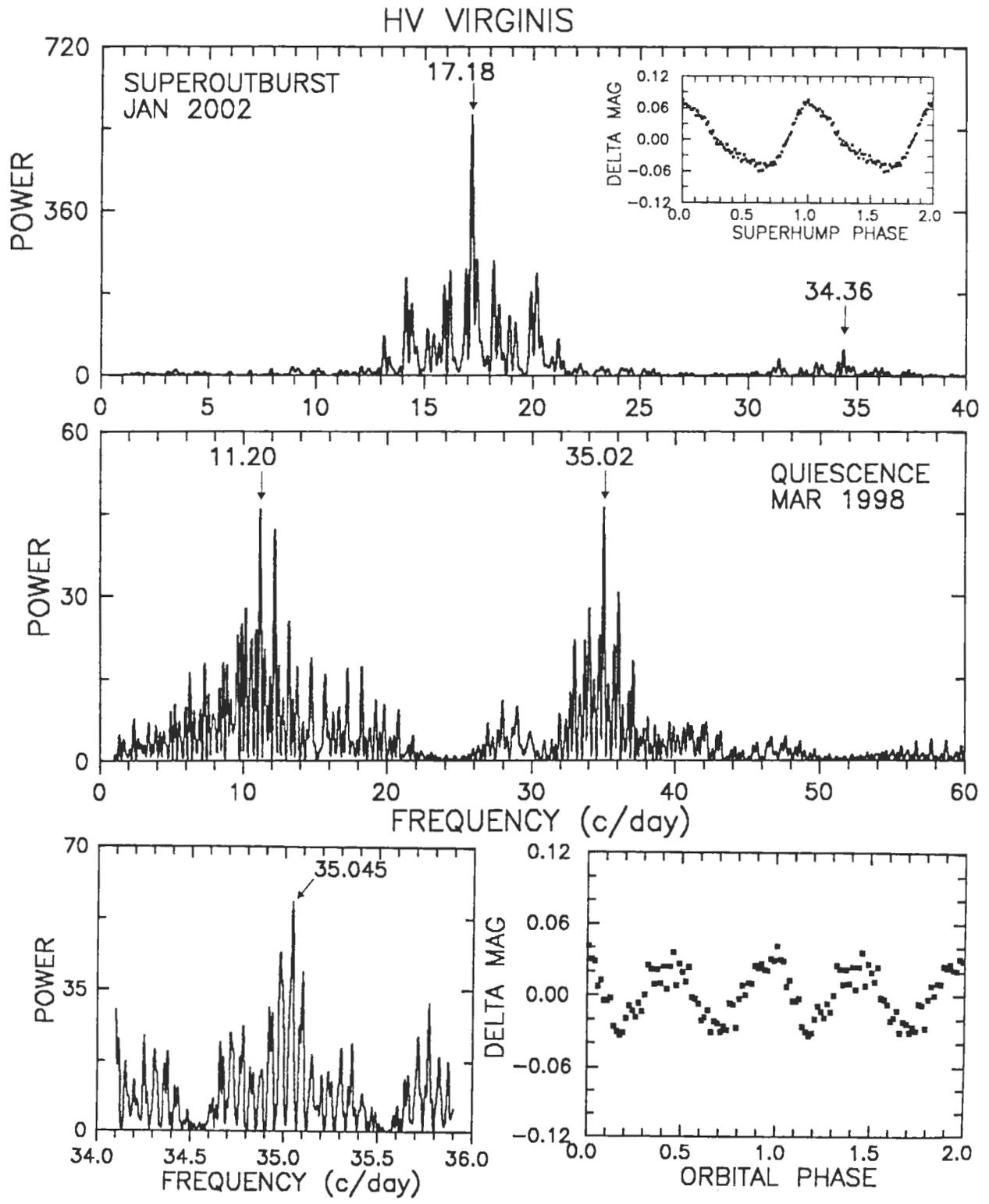

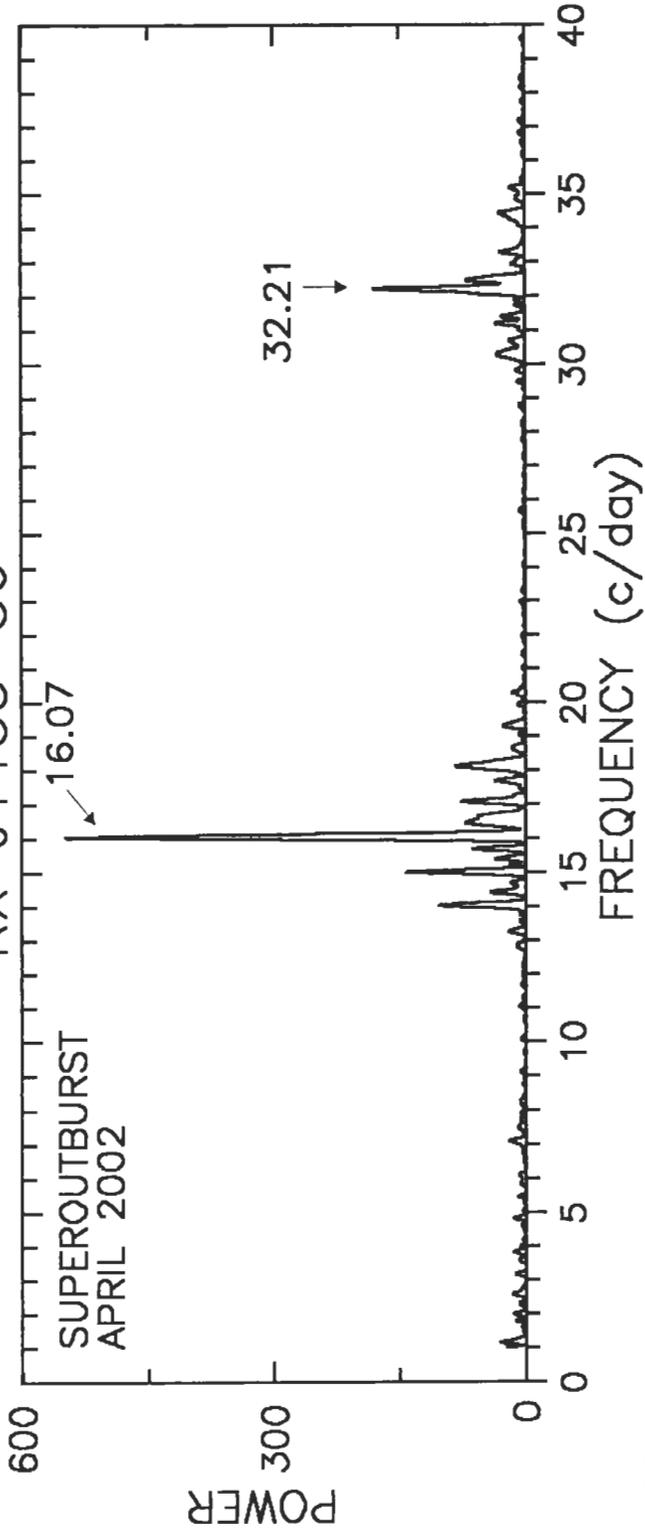
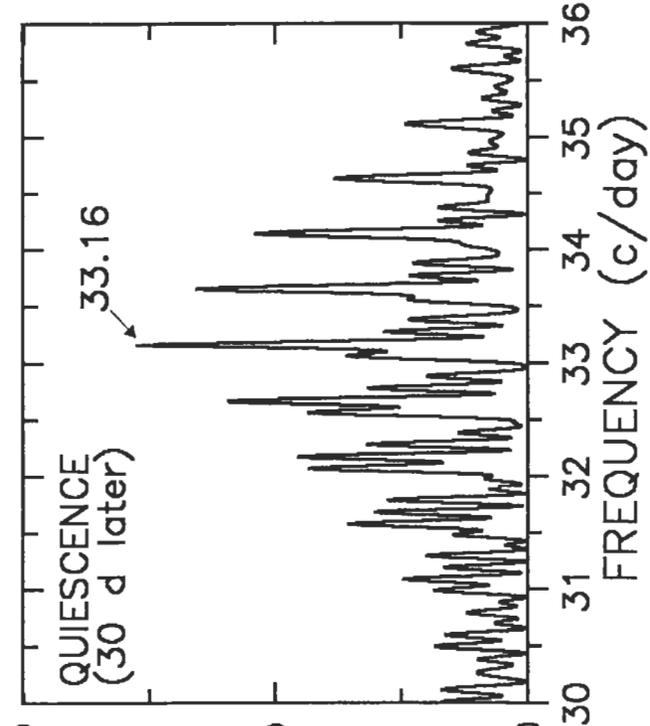
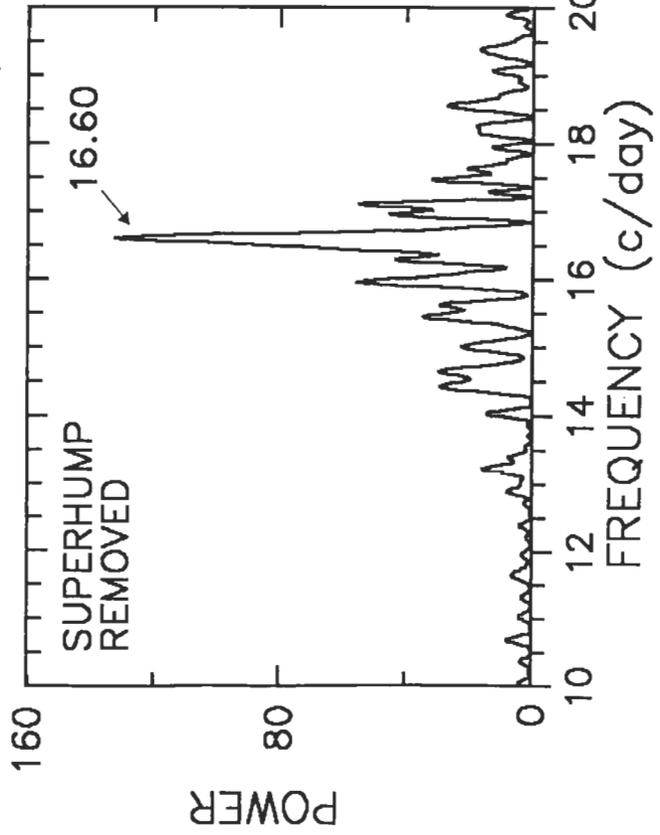

Fig 19

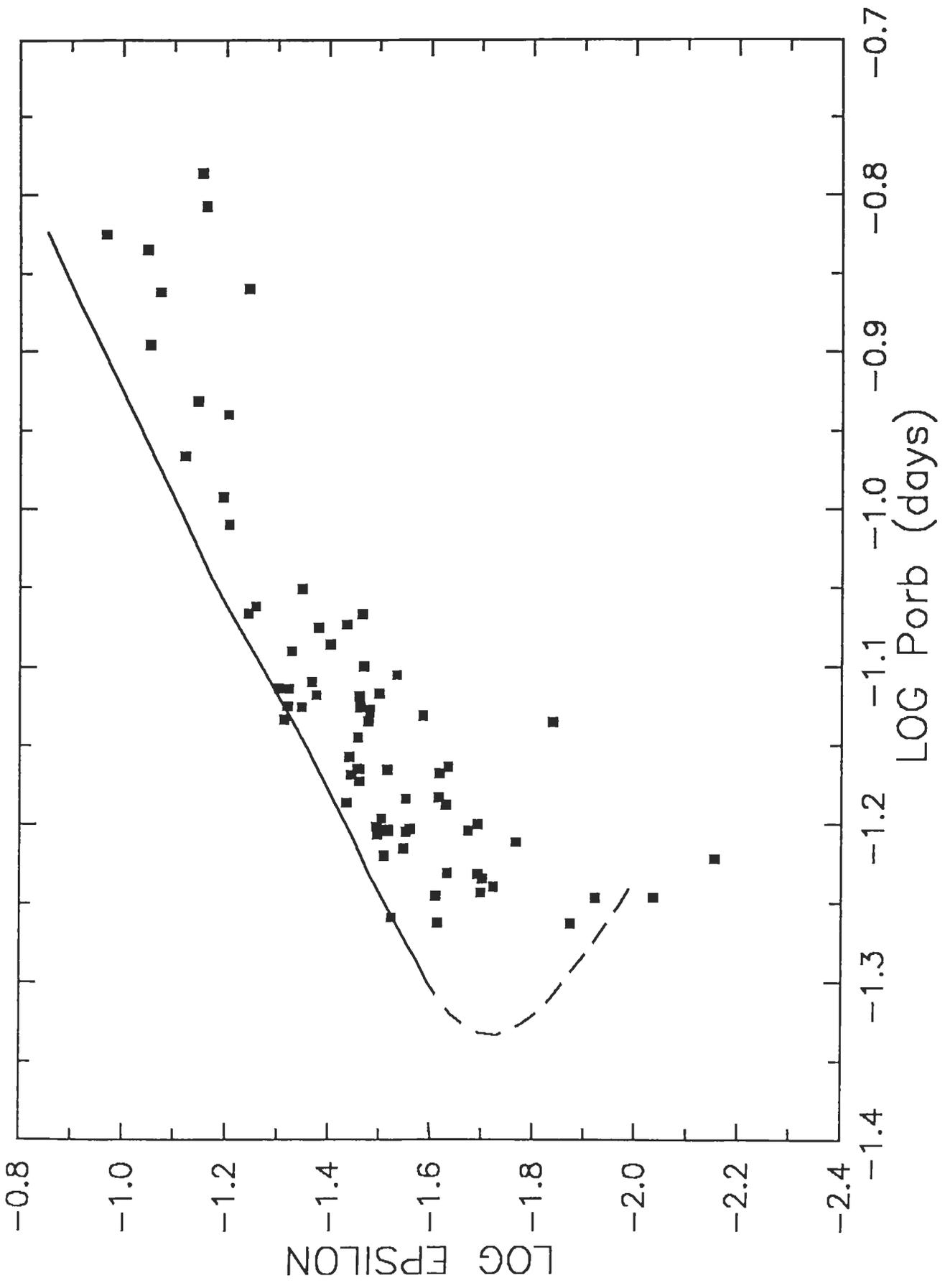

Fig 20

Fig 21

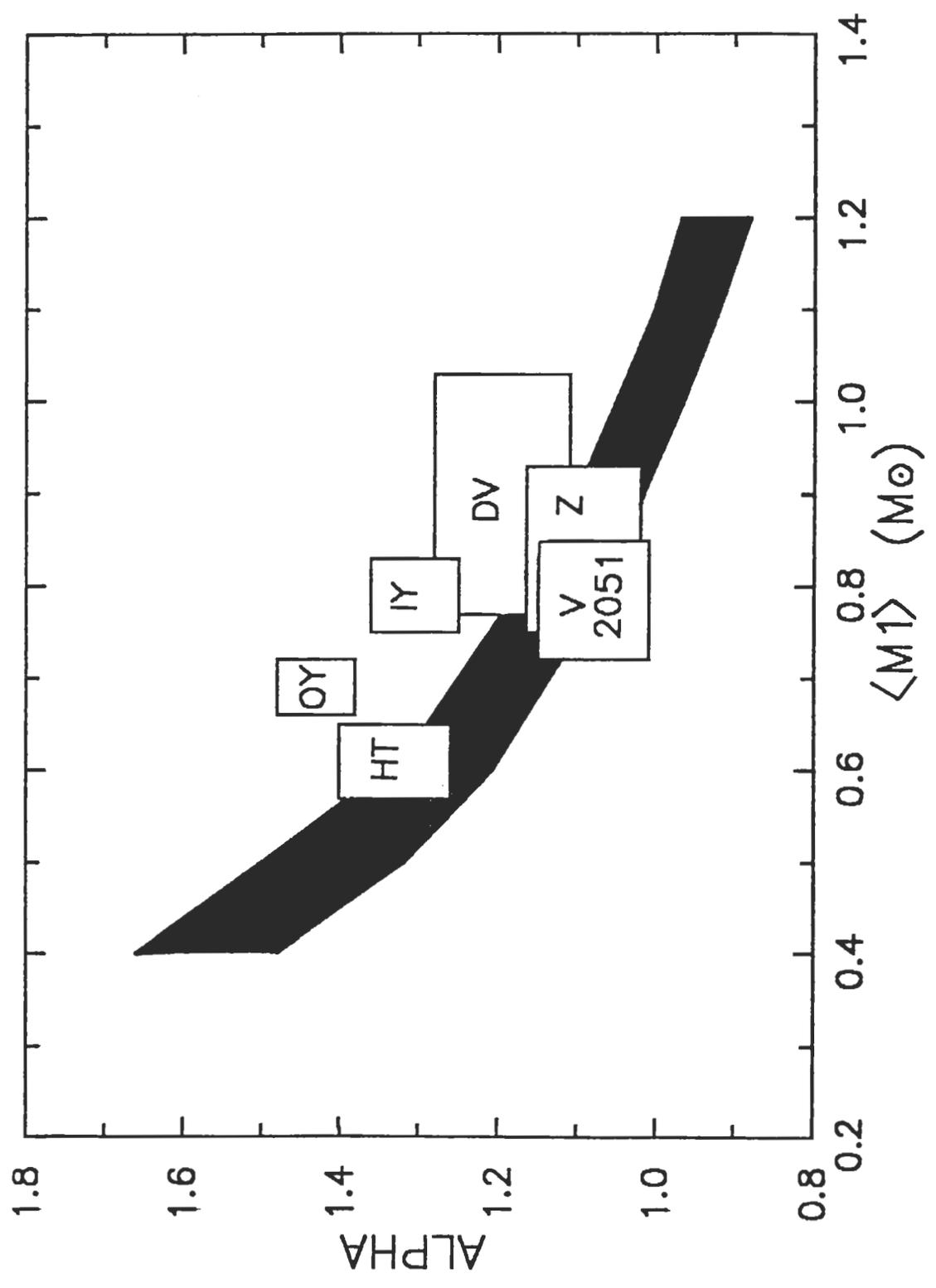